\documentclass{aastex}

\def\My{\hbox{\kern 0.20em My}}
\def\kms{\hbox{\kern 0.20em km\kern 0.20em s$^{-1}$}}
\def\cmmt{\hbox{\kern 0.20em cm$^{-3}$}}

\received{}
\accepted{}

\slugcomment{to appear in AJ}

\lefthead{Rosado et al.}
\righthead{Kinematics of ESO~379-006}

\begin{document}

\title{2D kinematics of the edge-on spiral galaxy ESO~379-006}

\author{M. Rosado\altaffilmark{1}, R. F. Gabbasov\altaffilmark{1}, P. Repetto\altaffilmark{1}, I. Fuentes-Carrera\altaffilmark{2}, P. Amram\altaffilmark{3}, M. Martos\altaffilmark{1},   and O. Hernandez\altaffilmark{3, 4}  }
\email{E-mail: margarit@astro.unam.mx}

\altaffiltext{1}{Instituto de Astronom\'\i a,
Universidad Nacional Aut\'onoma de M\'exico, Apartado Postal 70-264, CP 04510, M\'exico, D. F., M\'exico.}
\altaffiltext{2}{Escuela Superior de F\'isica y Matem\'aticas, Instituto Polit\'ecnico Nacional, U.P. Adolfo L\'opez Mateos, Zacatenco, 07730, M\'exico, D. F., M\'exico.}
\altaffiltext{3}{Laboratoire d'Astrophysique de Marseille, Aix-Marseille University \& CNRS, 38 rue Fr\'ed\'eric Joliot-Curie, 13338 Marseille CEDEX 13, France}
\altaffiltext{4}{ D\'epartement de Physique and Observatoire du Mont M\'egantic, Universit\'e de Montr\'eal, C.P. 6128, Succ. centre-ville, Montr\'eal, Qc, Canada H3C 3J7}

\begin{abstract}

We present a kinematical study of the nearly edge-on galaxy ESO~379--006 that shows the existence of extraplanar ionized gas. With Fabry-Perot spectroscopy at H$\alpha$, we study the kinematics of ESO~379--006 using velocity maps and position-velocity diagrams parallel to the major and to the minor axis of the galaxy. We build the rotation curve of the disk and discuss the role of projection effects due to the fact of viewing this galaxy nearly edge-on. The twisting of the isovelocities in the radial velocity field of the disk of ESO~379--006 as well as the kinematical asymmetries found in some position-velocity diagrams parallel to the minor axis of the galaxy suggest the existence of  deviations to  circular motions in the disk that can be modeled and explained with the inclusion of a radial inflow probably generated by a bar or by spiral arms. 
 We succeeded in detecting extraplanar Diffuse Ionized Gas in this galaxy. At the same time, from the analysis of position-velocity diagrams, we found some evidence that the extraplanar gas could lag in rotation velocity with respect to the midplane rotation.  
\end{abstract}


\keywords{Galaxies: kinematics and dynamics -- Galaxies: structure -- Galaxies: spiral-- Galaxies: ISM -- ISM: kinematics and dynamics -- Galaxies: Individual: ESO~379--006
}

\section{Introduction}

Some spiral galaxies have, in addition to ionized gas distributed in a thin disk whose width is similar to that of the stellar disk, another gas component revealed as a thick disk or halo. This component, now called Diffuse Ionized Gas, DIG, was identified for the first time in our own Galaxy by Ron Reynolds and collaborators using Fabry-Perot interferometers (Reynolds Layer; Reynolds, Roesler \& Scherb,1973). Now, by means of
HI and H$\alpha$ surveys of the Milky Way, together with the  synchrotron Galactic background,
Faraday rotation measures and cosmic rays, it is common knowledge that in our
Galaxy coexist several gas components:  a cool, thin disk of gas linked to the stellar disk, another thick, highly
pressurized layer of both cool and warm disks of different scaleheights
(Lockman 1984, Kutyrev \& Reynolds 1989), as well as a hot halo component extending
to substantial scaleheights from the midplane (Kalberla et al. 1998).
 The warm ionized component or DIG, has scaleheights of the order of 1 kpc (Madsen, Reynolds \& Haffner, 2006) while there has been also detected an HI layer with scaleheight of 400 pc (Lockman 1984). 

In a complementary approach, it has been found that in galaxies viewed edge-on, mostly late-type spirals having relatively high rates of star formation, layers of DIG and HI are detected at several kpc  above or below the galactic plane coining the name of extraplanar DIG or e-DIG for the diffuse ionized gas outside the disk of a galaxy and reaching vertical heights up to 8--10 kpc (Rand, Kulkarni \& Hester 1990, Rand 1996, Rossa \& Dettmar 2000, 2003a and 2003b, Matthews \& de Grijs 2004). 

The archetypical galaxy where these studies started is the nearby Sbc spiral NGC~891 (Rand, Kulkarni \& Hester 1990). In this galaxy e-DIG  has been detected  for Z-heights up to 5 kpc (Rand 1997, Hoopes et al. 1999) and it has been found that the [SII]/H$\alpha$ line-ratio of e-DIG was higher than for classical HII regions (Otte, Gallagher \& Reynolds, 2002). Neutral gas has also been detected in NGC 891 well beyond Z-heights of 5 kpc (Oosterloo, Fraternali \& Sancisi 2007) and the HI kinematics seem to imply that the emission is due to a HI
halo lagging by 25-100 km s$^{-1}$ in rotation speed  relative to the gas moving
in the midplane. The lagging of the halo gas has also been measured for the ionized gas or e-DIG by Heald et al. (2006a and b). 

For galaxies not exactly seen edge-on, DIG and e-DIG could be present as well but their detection is complicated because projection effects preclude to distinguish it from HII regions in the galactic plane. Whether the DIG and the e-DIG phases are similar and formed by the same mechanisms is a question of debate and still open, at least in what concerns their kinematical behavior.
It is thus important to know if that gas component, not directly associated with stars, has some distinctive properties which could be used to identify it even for galaxies seen at other inclinations. In that respect 
follow-up studies of DIG and e-DIG seem to show that some emission line-ratios of the DIG component such as [SII]/H$\alpha$, [NII]/H$\alpha$ or [OIII]/H$\beta$ have larger values than those expected from classical photoionized HII regions (Haffner, Reynolds, \& Tufte 1999; Golla, Dettmar, \& Domgorgen 1996; Rand 1998).

The observations of this ionized gas in the case of our Galaxy allow us to know more about its physical conditions: very low density (0.1 cm$^{-3}$), warmer electron temperatures  (10$^4$ K) in about 2000 K than those of classical HII regions, and [SII]/H$\alpha$ and [NII]/H$\alpha$ line-ratios much higher than the characteristic of typical HII regions (Madsen, Reynolds \& Haffner 2006).

Some edge-on galaxies have their entire disks encompassed in a thick (about a few kpc thick) layer (halo or thick disk) of ionized gas. That is the case for NGC~891. Other edge-on galaxies have only local signs of e-DIG. For these later galaxies, instead of the pervasive layer of diffuse gas encompassing the disk, the e-DIG morphology shows a wide variety of local features such as plumes and filaments (mostly perpendicular to the disk) concentrated in specific regions above or below the midplane, as the direct imaging observations of Rossa \& Dettmar (2000, 2003a) show. However, so far there is no evidence of any smooth, diffuse gas encompassing those features.

In what concerns the DIG kinematics, most of the work comes from edge-on galaxies where there are some signs of vertical gradients of rotation velocity or simply a lag in rotation velocity relative to the disk velocity (Kamphuis et al. 2007, Heald et al. 2006a and Heald et al. 2006b). But still remains to be studied whether the velocity dispersions of the DIG are larger than those of the gas in classical HII regions as some works appear to suggest (Valdez-Guti\'errez et al. 2002). Also, in the case of non edge-on galaxies (see for  instance the study by Fraternali et al. 2001 and 2002 on the spiral galaxy NGC~2403) it has been possible  to discriminate the gaseous halo against the projected disk gas of the galaxies by kinematical means. 
It was becoming clear that the DIG component shows different kinematic properties than those of the ionized gas linked to massive stars and distributed in the thin galactic disk (Fraternali et al. 2001 and 2002, Matthews \& Wood 2003, Uson \& Matthews 2003, amongst others). Indeed, for the galaxies where thick disks  have been identified and studied kinematically (both edge-on and non edge-on), it has been shown that the thick disk component does not follow the general rotation of the galaxy but it is lagging in rotation by 20 to 100 km s$^{-1}$. Furthermore, the kinematics of the thick disk component seem to be a function of the height above or below the disk (coordinate Z in a cylindrical system centered at the galactic center). For the ionized gas (e-DIG) the kinematics is better studied with 2D spectroscopy, in an analogous way as the HI kinematics.

These latter points can shed light on the origin of the DIG component. The main process invoked to explain the e-DIG is the  galactic fountain mechanism (Shapiro \& Field 1976).  However, it is possible that other processes could concur besides the SF mechanism. In particular, we should take into account that the presence of a bar or of spiral arms, taking those entities as obstacles, induces kicks (known as hydraulic jumps) of gas from the disk to the halo that could be linked to  the presence of some e-DIG  in barred galaxies. A bar could also induce radial inflow/outflow motions which could be important till 2.5 times the bar length (Martos, Priv. Comm.). 

Aiming at unraveling some of the issues discussed above, we planned to study
the 2D kinematics of the e-DIG of a galaxy where e-DIG was already detected. In this work we present our H$\alpha$ scanning Fabry-Perot observations on the galaxy ESO~379-006 and study its 2D kinematics. A confrontation of the kinematical data with some galaxy modeling allows us to explain several of the kinematical peculiarities of this galaxy.

ESO~379-006 is an almost edge-on  galaxy catalogued with that number in the ESO-Uppsala Catalog of Galaxies. Extraplanar DIG was detected in this galaxy up to heights of 2 kpc by Rossa \& Dettmar (2003a and 2003b), in spite this galaxy has not an IRAS FIR flux ratio S$_{60}$/S$_{100}$  $\ge$ 0.4, which seems to be a criterion  to grant the detection of e-DIG in late type spirals. Rossa \& Dettmar (2003a and 2003b) find for this galaxy an e-DIG morphology of the type of locally extended emission and plumes. 
These authors derive, from IRAS observations on this galaxy, an star formation rate, SFR of 1.49 M$_\odot$ yr$^{-1}$ (comparable to the one derived for NGC~891), a frequency for supernova explosions, $\nu$$_{SN}$ = 0.061 yr$^{-1}$, and an energy input  from supernova explosions per unit area of star formation activity (a kind of SN energy flux), dE$_{SN}$/dt $\times$ Area$_{SF}$$^{-1}$ = 4.1 $\times$ 10$^{-4}$ erg s$^{-1}$ cm$^{-2}$, i.e., about 10 times smaller than the one derived for NGC 891. 
Table 1 gives the main characteristics of ESO 379-006.

Results on the kinematics of this late-type galaxy  reported by Mathewson, Ford \& Buchhorn (1992), obtained from H$\alpha$  and [NII] slit spectroscopy, are given in Table 2.
Table 2 also quotes kinematical and e-DIG properties derived from observations of Mathewson et al. (1992), Persic \& Salucci (1995), Rossa \& Dettmar (2003a and 2003b) and this work. As seen in Table 2, the value adopted for the inclination of this galaxy varies from 81$\arcdeg$ to 90$\arcdeg$.

The paper is organized as follows:  Section 2 describes the observations and data reduction,  Section 3 presents the results on the morphology of this galaxy obtained from the imaging. Section 4 presents the kinematical data (velocity maps or velocity channels, position-velocity diagrams, radial velocity fields, non-circular velocity fields) and their confrontation with the predictions of simple galaxy modeling varying different kinematical components. In Section 5, a general discussion and the conclusions are laid-down.

\section{Observations and data reduction}

The observations of the galaxy ESO~379-006 were conducted at the ESO 3.6m telescope during the night of April 26, 2004. The CIGALE instrument was attached to the Cassegrain focus of the telescope in order to provide Fabry-Perot (FP)  data cubes of the galaxy. The data were registered with an IPCS detector (a 2D photon counting system of 512 $\times$ 512 pixels).

The CIGALE instrument  provides data cubes with a field of view of 3$\arcmin$.5 and a  scale of 0$\arcsec$.405 pix$^{-1}$. The high order, scanning FP  was an ET-50 from Queensgate Instruments, Ltd. with a ``Finesse'' of 18 and an interference order of 609 at H$\alpha$ at rest. In order to satisfy Shanon's rule, 40 different steps (channels) were obtained, regularly spaced over the FP's free spectral range of 493 km s$^{-1}$ (10.89 \AA),  providing @ H$\alpha$, a spectral resolution R $\sim10850$, a FWHM of spectral PSF of $\sim27.4\kms\sim0.605\AA$ and a spectral sampling of $\sim12.3\kms\sim0.27\AA$. Taking advantage of the use of a photon-counting detector, the elementary exposure time per scanning step was 15 seconds. Thus, each cycle of 40 channels has been integrated during 40 $\times$ 15 sec and, 16 cycles were obtained for the galaxy, implying a total exposure time of 2.67 hours. We used an interference filter centered at $\lambda$6630 \AA~ and with a FWHM of 20 \AA~ to isolate the redshifted H$\alpha$ emission of the galaxy, ensuring that the full H$\alpha$ line of the galaxy is detected.

In photometric studies, flat-fielding is critical since absolute flux measurements strongly depend on the transmission across the field of view. In kinematic studies, as far as the flat field variations are low frequency (which is the case here) and the measurements on each profile done locally for each pixel  are higher moment measurements on the emission lines above the continuum level (center, line width, skewness, kurtosis), they do not depend on the flat field correction. In view of this, and based on previous experiences with this instrument, we did not have performed any flat field correction to avoid adding noise.

 The calibration of the data cubes was carried out by obtaining, under the same observing conditions, calibration cubes using the line at $\lambda$6598.95 \AA~ of a diffuse Neon lamp. The calibration cubes were obtained before and after the object observations in order to check for possible flexures of the instrument. 
In this way, a calibrated object datacube was produced.

Reduction of the datacubes was carried out using the CIGALE/ADHOCw software (Boulesteix 1993). The data reduction procedure has been extensively described in Fuentes-Carrera et al. (2004) and references therein. The accuracy of the zero point for the wavelength calibration is a fraction of a channel width ($\leq$ 3 km s$^{-1}$) over the whole field. OH night-sky lines passing through the filter were subtracted, using the same datacube of the galaxy, by determining the emission in several regions outside the galaxy (but nearby to it) taking advantage that the field of view of the instrument is much larger than the region covered by the galaxy. In this way, we are sure of subtracting a sky contamination present (which is highly variable with time) under the same conditions than the one over the galaxy. 

With the ADHOCw software we have done a series of tasks: first, to calibrate in wavelength the original FP interferogram data cubes producing a data cube formed of 40 channels or velocity maps corresponding to very narrow interference filter images at the same radial velocity (FWHM of 0.27~ \AA) suitable to detect extended emission features such as the ones searched for the e-DIG; to smooth the velocity cubes either spatially or spectrally; to extract radial velocity profiles pixel per pixel (or integrated over more larger zones);  and to extract continuum-free, integrated line intensity and continuum images obtained by integrating the intensities of the radial velocity profiles up or down a certain percentage of the peak value, respectively. The continuum images, allowed us to determine the photometric center of the galaxy. The data cube and the continuum-free, integrated line intensity images or H$\alpha$ images are useful to detect DIG emission. In this particular case, in order to enhance the S/N ratio, a Gaussian spatial smoothing was done ($\sigma$ = 3 pixels) in the data cube. The corresponding angular resolution of the smoothed cubes is 1$\arcsec$.2 (218 pc). The measured intensities of lines are in units of signal-to-noise ratio since we do not obtain absolute flux calibration. The estimated noise of these cubes is about 7 counts on each channel. 
ADHOCw also allowed us to extract from a data cube two or more cubes obtained from the fitting of two or more Gaussian velocity components to the line of sight velocity profiles, or to subtract one of those  cubes from the observed one in order to remove some spurious component.

We also used IRAF {\footnote{ IRAF is distributed by the National Optical Astronomy Observatories, operated by the Association of Universities for Research in Astronomy, Inc., under cooperative agreement with the National Science Foundation.}}  tasks, Python routines to obtain and plot position-velocity diagrams and the tasks MOMENT, GALMOD and VELFI from the GIPSY package (distributed publicly by the Kaptein Institute) to fit galaxy models to the observed velocity data.

\altaffiltext{7}{ IRAF is distributed by the National Optical Astronomy Observatories, operated by the Association of Universities for Research in Astronomy, Inc., under cooperative agreement with the National Science Foundation.}


\begin{figure}
\includegraphics[angle=0,width=\columnwidth]{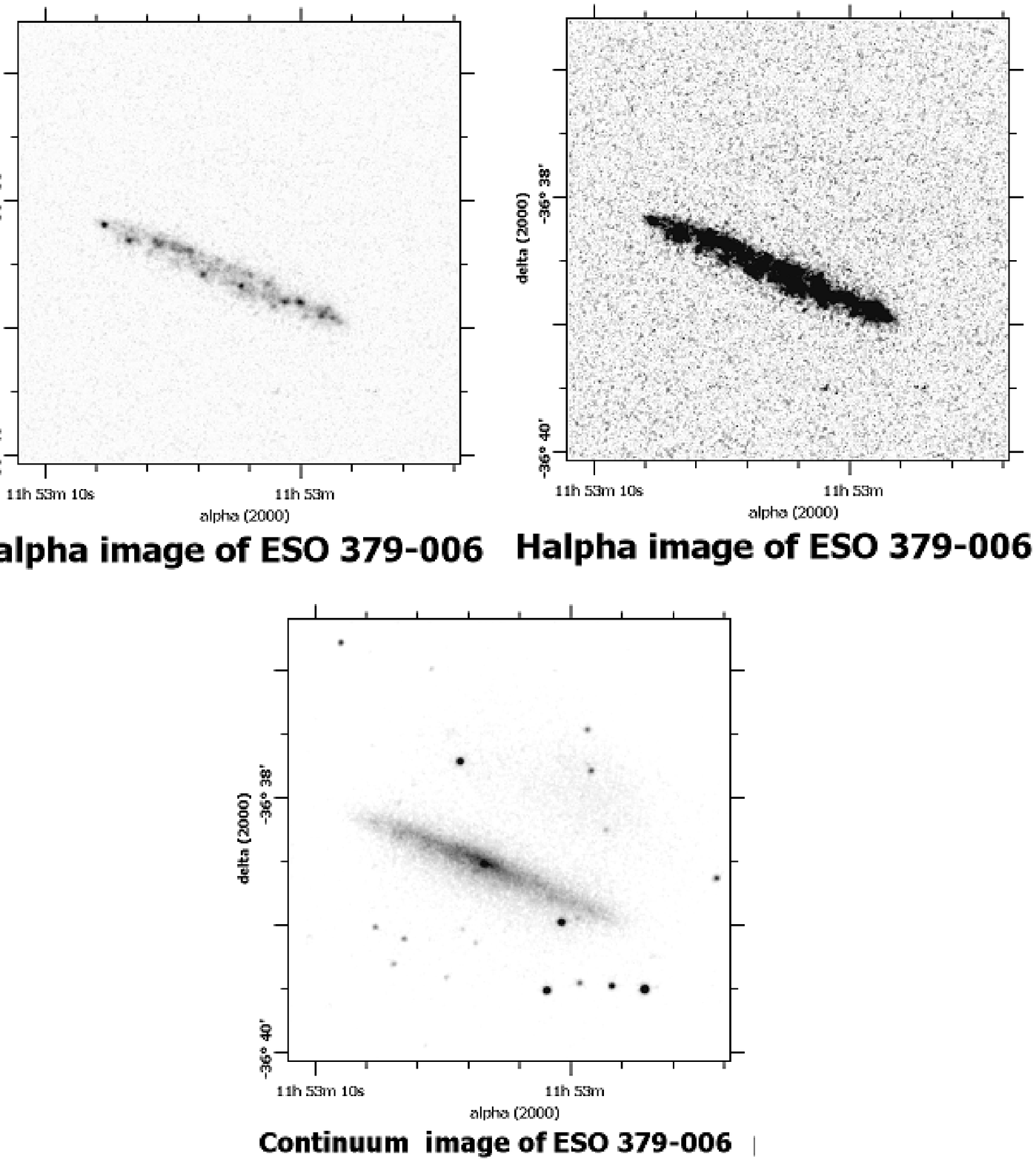}
\vskip 0.5in
\figcaption{\label{fig. 1} H$\alpha$ images (high contrast: top left, low contrast: top right) and continuum image (bottom) of the galaxy ESO~379-006 obtained from the FP data cubes as discussed in Section 2. The NW patch of emission in the continuum image is due to internal reflections in the instrument.}
\end{figure}


\section{H$\alpha$ imaging  of ESO~379-006}

We proceeded to search for the e-DIG of ESO~379-006 starting with the  H$\alpha$ images. Figure 1 shows the  H$\alpha$  and the continuum emission images obtained from our FP data cube, according to the procedure described in Section 2. In this figure, the H$\alpha$ image is shown at two different contrast levels allowing to note the bright midplane HII complexes  and the more tenuous filaments presumably corresponding to e-DIG, respectively.  The left panel H$\alpha$ image shows the  distribution of bright HII complexes in the disk. It appears in this image that ESO ~379-006 is not quite edge-on because the HII regions seem to form an elliptical ring (see Figure 1; this ring is not seen in the continuum image) of high ellipticity but not the straight line expected for a galaxy really viewed edge-on.
Considering the major and minor axis reported in the NED for the continuum emission, we will adopt a value for the inclination ($\it i$ = cos$^{-1} (b/a)$)  of 82$\arcdeg$.5.
 It is not excluded that the HII regions are not located on a circular ring but belong to spiral arms; in that case, depending on the unknown pitch angle and orientation of the spiral arms, the inclination value could be slightly higher or lower than the adopted value.

In order to know which side of the galaxy is the nearest to us we proceeded in two different ways: First we compared the relative intensities of the HII complexes located in both sides of the galaxy. Although not flux calibrated, our H$\alpha$  image allowed us to estimate the relative intensities of the HII complexes situated along the elliptical ring mentioned above (see the H$\alpha$ image in the left panel of Figure 1). It results that the HII complexes located in the southern side of the galaxy are brighter than those located in the northern side, suggesting that the southern side is the closest to us, although it could be possible that the HII complexes of the southern side are intrinsically brighter than those in the northern side. Second, we obtained an intensity profile from a cut of the continuum image (also displayed in Figure~1) along the minor axis of ESO~379--006. According with de Vaucouleurs (1958) and Sharp \& Keel (1985), authors who take into account extinction effects, the surface brightness profile along the minor axis of the near side of the disk of a galaxy falls more abruptly than the profile along the far side. We found that the intensity profile corresponding to the southern side falls more abruptly than the intensity profile corresponding to the northern side, thus in agreement with our previous result based on the HII region brightness. Consequently, for ESO~379--006, the southern side is the closest to us.

Since this galaxy is not exactly edge-on, care should be taken in order to estimate whether some e-DIG is detected or instead, it is only the projected disk gas, or a combination of both but without any clear distinction between the disk gas and the e-DIG. The extension of the ring of HII regions is, according to our image, about 128$\arcsec$ (equivalent to 23.3 kpc); by assuming that it is circular, and that the disk is infinitely thin, the height of the projected disk would be: $\it a$ $\times$ cos$\it i$, i.e., $\pm$8$\arcsec$.5 (equivalent to 1.54 kpc, where $\it a$ is half the extension of the ring and $\it i$ is the inclination of the galaxy). In Figure 2   we display the  H$\alpha$ image with overlayed contours of equal surface brightness of bright HII regions in the disk of the galaxy. We also overlay a geometrical sketch considering the e-DIG distributed in a cylinder extending up and down the galactic plane. The image, contours and geometrical sketch have been rotated in order to have the major axis coincident with the X-axis. It is possible that the disk is not circular in which case, some of the gas outside the circle may belong to the disk, i.e., to an external spiral arm.

In order to check this possibility and to help in the e-DIG identification from our images, we required to discriminate the gas emission in the disk from the gas emission outside the disk seen in projection. Since the disk is better appreciated from its stellar emission, we  superimposed our H$\alpha$ image with the contours of the continuum image. Thus, the stars revealed by the continuum image belong to the disk and consequently, the continuum  contours depict the galactic plane. Consequently, any emission outside the galactic plane could be considered as e-DIG. Figure 3 shows this superposition together with some regions of H$\alpha$ emission outside the continuum contours (marked with arrows) which could be considered as e-DIG. In this case, the FP cube used to produce Figure 3, the data cube and the radial velocity profiles have been smoothed spatially with a Gaussian function of $\sigma$ = 5 pixels (corresponding angular resolution of 2$\arcsec$ or about 363 pc) better suited to unravel faint filamentary features.
It is necessary to note that we cannot decrease the contour levels in the continuum image much more than the level shown in Figure 3 because we have found that for this galaxy the axis ratio increases with the distance to the center giving even lower values for the inclination that are not supported by the models we have run below (see Sect. 4.3.3). Indeed, the axis-ratio of the contours in Figure 3 implies an inclination of about 80$\arcdeg$, already lower than the adopted value. 

 We think that these places are e-DIG locations because in general,  that emission has  larger intensities than the noise peaks of the 2D H$\alpha$ image. Besides, in order to ensure that the emission is real, we have analyzed the radial velocity profiles of these features (obtained from the FP data cube; not shown) and we have found that they correspond to typical H$\alpha$ emission lines showing continuity in form among neighboring pixels, suggesting that the extraplanar features show similar velocities to the neighboring part of the disk, while the velocity profiles of marginally less brighter features corresponding to noise peaks in the 2D image are complex (mostly without a clear signal) and correspond to disconnected patches. In addition, the features  detected are much brighter than the noise in the individual velocity maps of the FP cube (each one in a different velocity map, not shown in this work).

Although those filaments are not distributed uniformly, there is no special concentration for the features are distributed at several different locations spreading over the extension of the galaxy. Some of the emission, the brightest and closest one, could be in the galactic plane if the disk is intrinsically elliptical instead of circular. However, some emission seems to be so far away from the midplane that their heights should amount to 1.84 kpc (Z$_{MAX}$ in Table 2, estimated from the angular distance to the major axis). The suggested e-DIG features have the appearance of filaments or tongues, some of them departing continuously from regions inside the projected disk height, forming coherent structures.


\begin{figure}
\includegraphics[angle=0,width=\columnwidth]{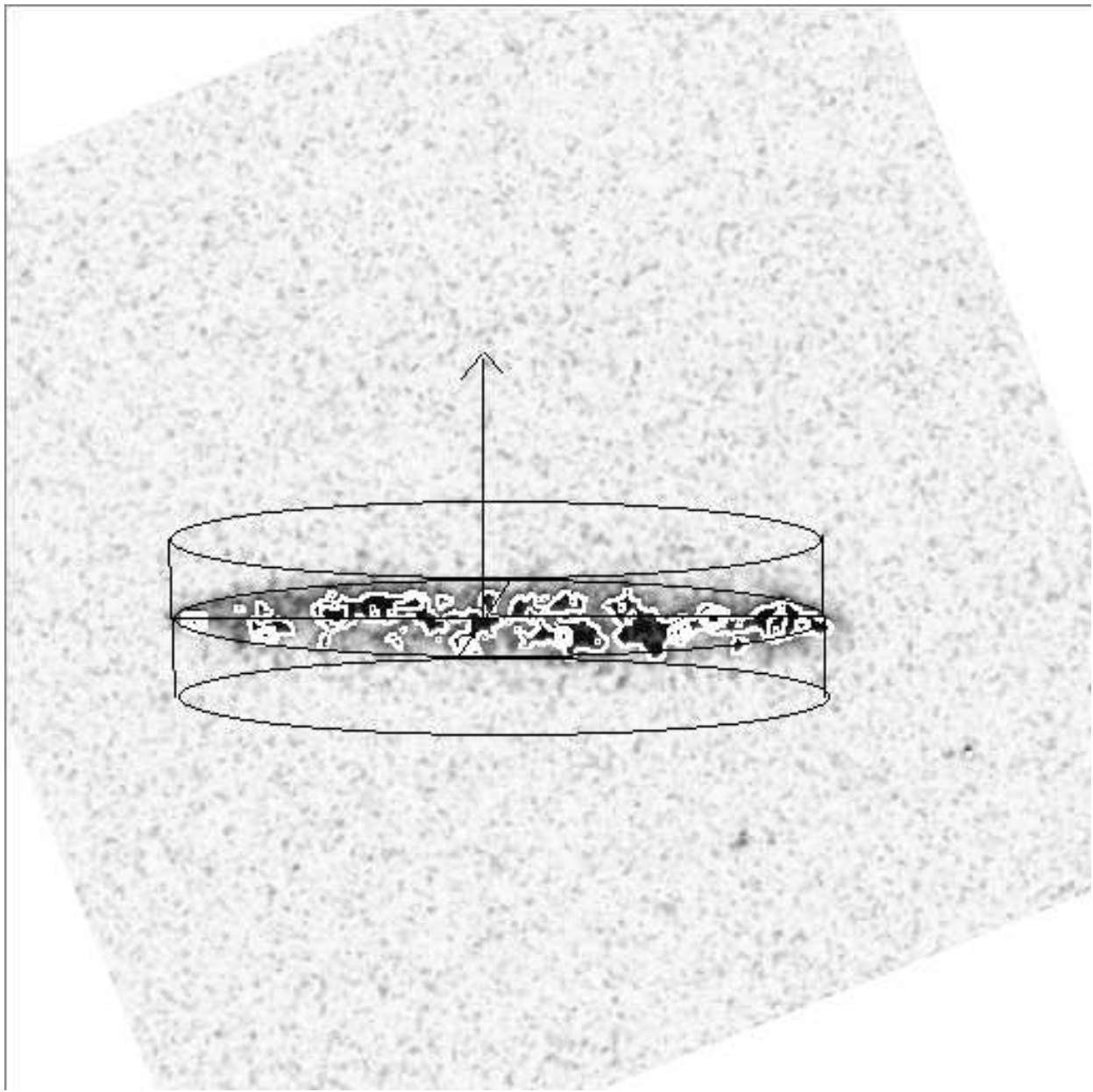}
\vskip 1in
\figcaption{\label{fig. 2} H$\alpha$ image of galaxy ESO~379-006 with overlayed contours of equal surface brightness. Also overlayed is a geometrical sketch of a cylindrical e-DIG distribution. The sketch was constructed for an inclination of about 82$\arcdeg$.5 and a cylinder's height above the disk plane of about 8$\arcsec$.5 (about 1.54 kpc). The image is rotated such as the major axis coincides with the X-axis. }
\end{figure}



\begin{figure}
\includegraphics[angle=0,width=\columnwidth]{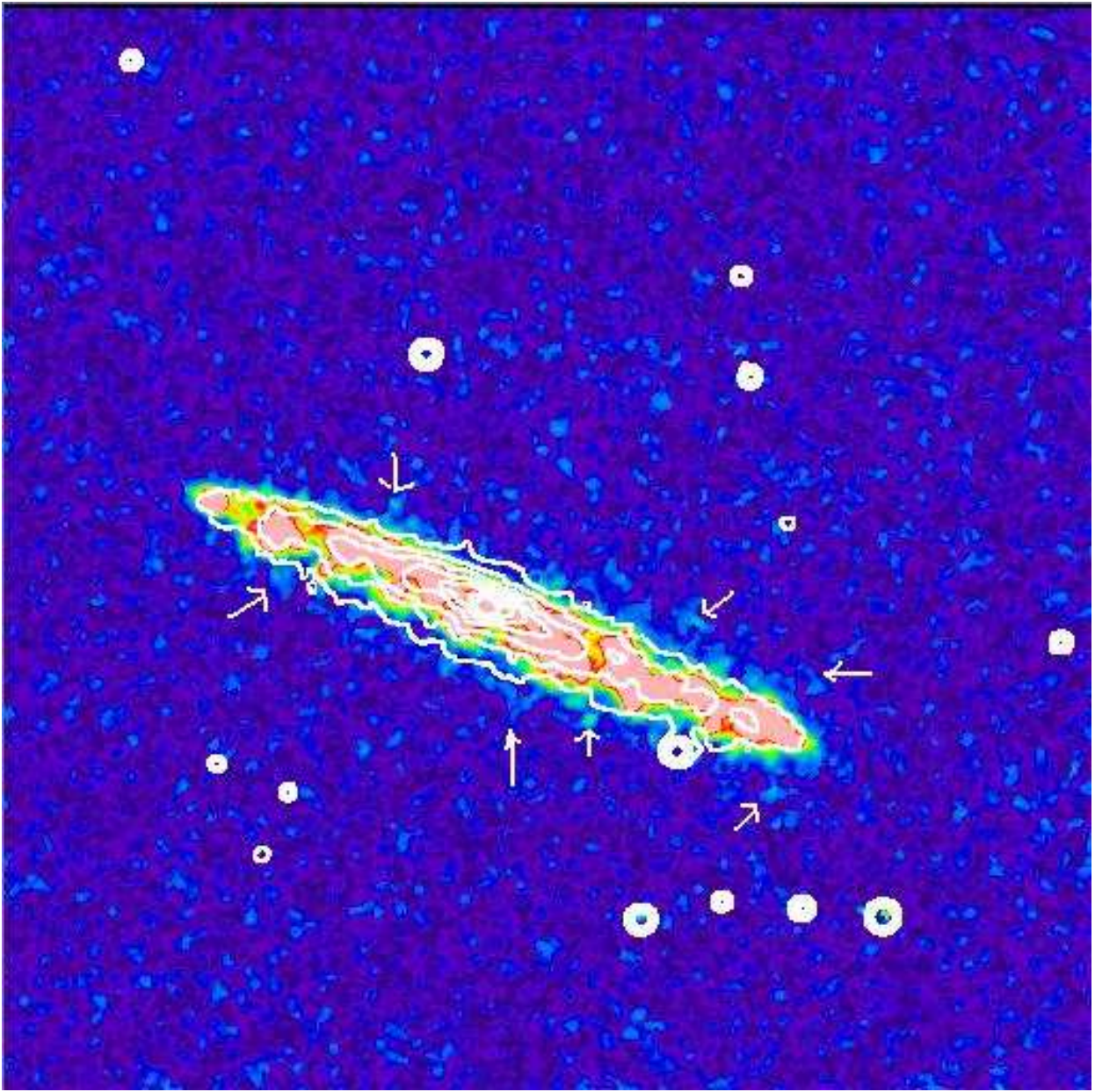}
\vskip 1.2in
\figcaption{ \label{fig. 3}H$\alpha$ image of ESO~379-006 with overlayed continuum contours revealing the stellar disk (white contours). The arrows mark possible e-DIG features (see the text for details). This image was Gaussian smoothed with $\sigma$ = 5 pixels (corresponding  resolution of 2 $\arcsec$ or 363 pc). }
\end{figure}



\begin{figure}
\includegraphics[angle=0,width=\columnwidth]{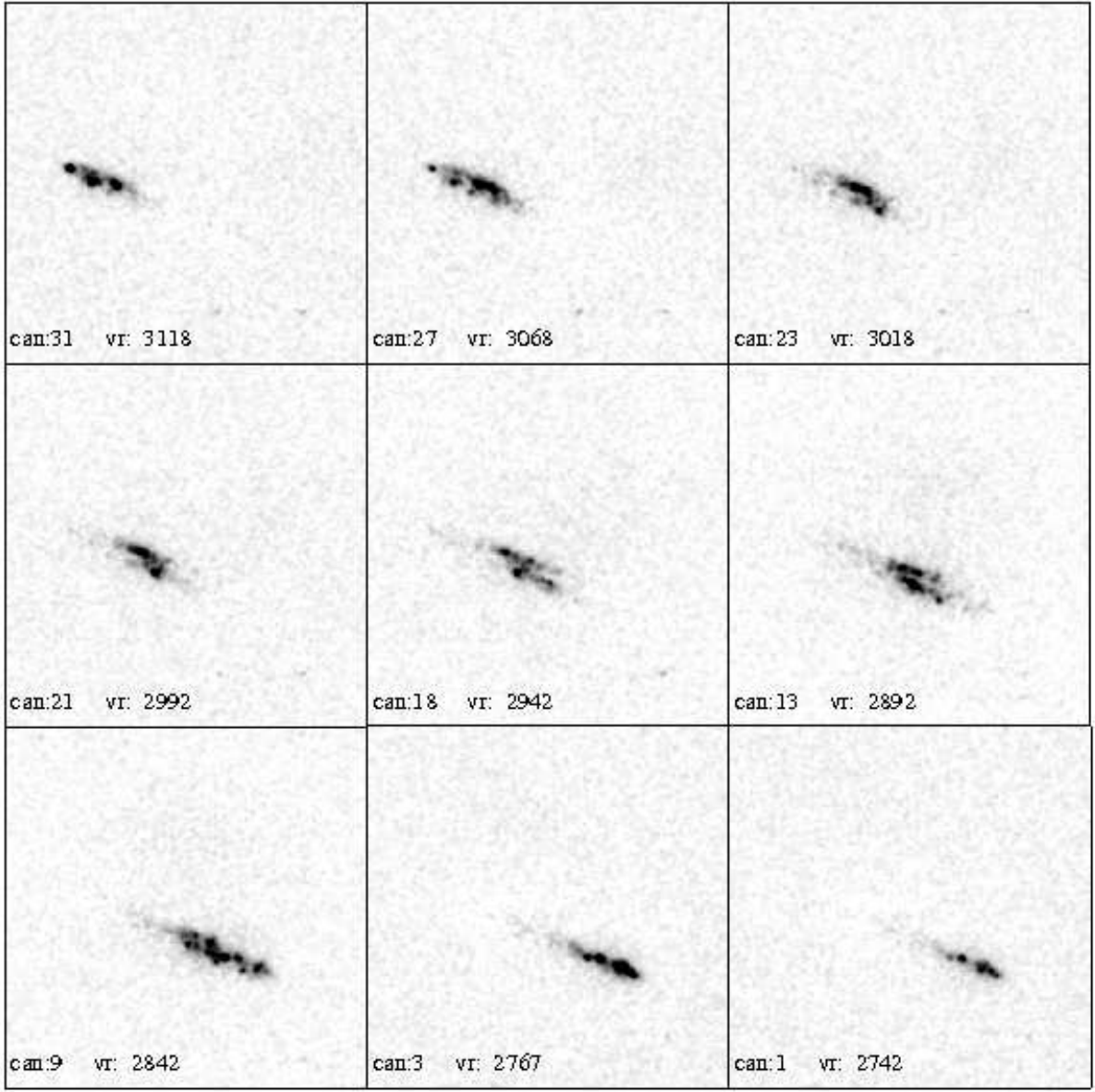}
\vskip 1.2in
\figcaption{ \label{fig. 4}Some of the H$\alpha$ radial velocity maps of ESO~379-006. The maps were obtained from our FP cube which was smoothed with a Gaussian function of $\sigma$ = 5 pixels (corresponding resolution of 2 $\arcsec$ or 363 pc). Note the bifurcation in channels appearing in the central panels.}
\end{figure}



\begin{figure}
\includegraphics[angle=0,width=\columnwidth]{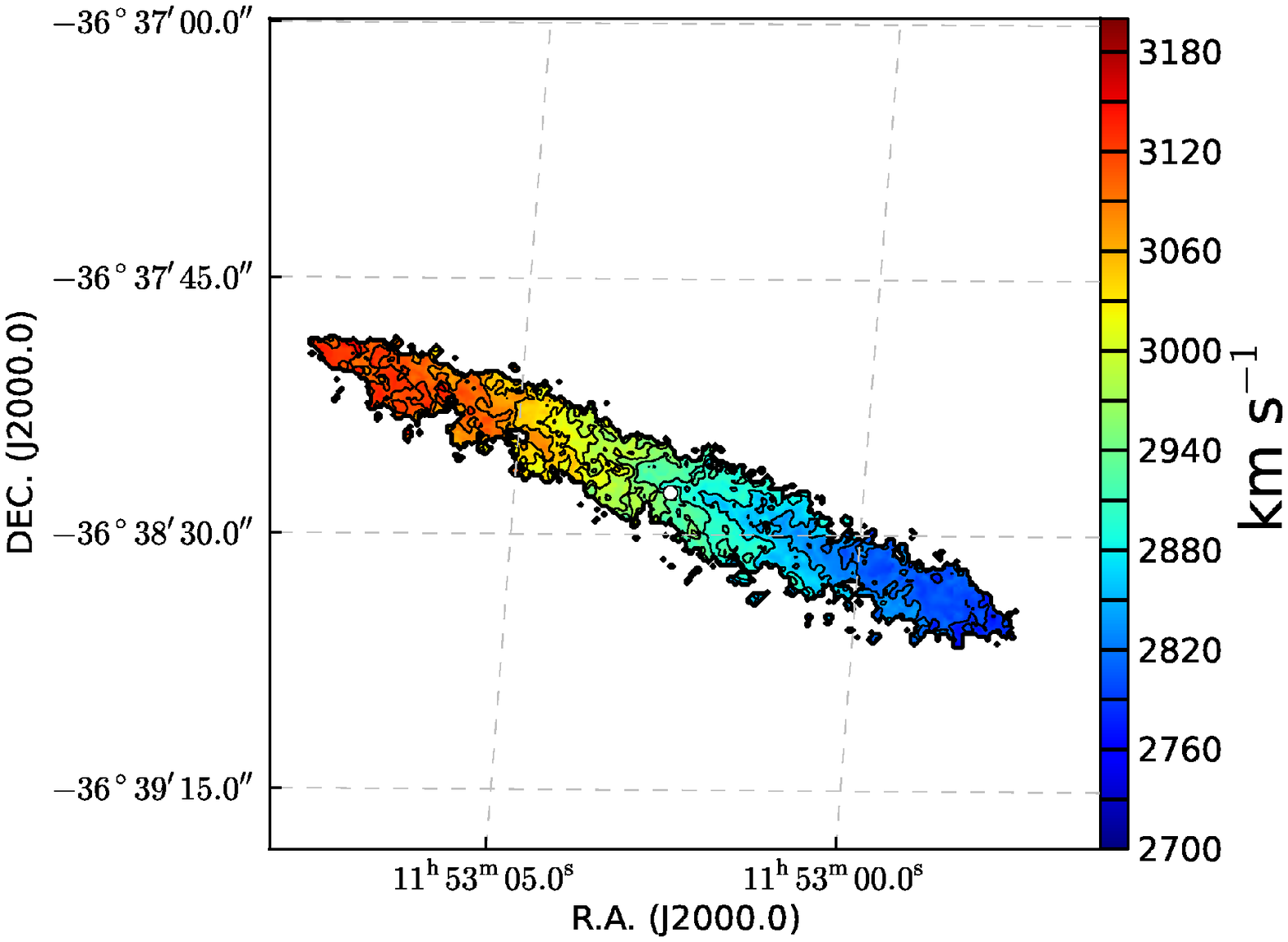}
\vskip 0.5in
\figcaption{ \label{fig. 5}Radial velocity field of the main component (disk) of ESO~379-006. Overlayed are equal velocity contours. The white dot stands for the photometric center. }
\end{figure}


\section{Kinematical Analysis of ESO~379-006}

\subsection{ Velocity maps, velocity field and velocity widths of the disk}

Figure 4  shows several velocity maps (or velocity channels) extracted from our FP datacube. In total, we have 40 velocity maps equally spaced, separated in 12.3 km s$^{-1}$. We can appreciate in this figure the bifurcations seen in several velocity maps which are relevant in determining the inclination of this galaxy (see Sect. 4.3.3). At other contrasts (not shown here) suitable to visualize fainter features we are able to follow several e-DIG features.

From our original data cube we have constructed the velocity field using the task MOMENT of the GIPSY
 package to produce the first moment map shown in Figure 5.
 In this figure equal velocity contours and the position of the photometric center are drawn. Some global trends can be identified:
 {\it i}) A rotation is clearly seen being the NE side receding and the SW side approaching.
 {\it ii}) The kinematic center (see Sect. 4.2.1 for its identification) does not coincides with the photometric center of the galaxy, but it is rather shifted in the north-east direction (see Table 2).
 {\it iii}) The kinematic minor axis is well aligned with the morphological minor axis.
 {\it iv}) The equal-velocity contours are skewed, suggesting the existence of non-circular motions.
We will analyze in more detail the twisting of the velocity contours and its possible causes in Section 4.3.
 
We have also constructed  the second moment map of our data cube which is shown in Figure 6. In this figure we note two strips parallel
 to the major axis with small velocity dispersions (low signal-to-noise regions tend tend to have broader lines than the high signal-to-noise strips due to a lack of flux to propoerly resolve the profile).
The strips are also seen in  Figures 1
 and  2 as a chain of bright H$\alpha$ spots. We think that it is an indication of possible enhanced spiral arms or ring, suggesting that the galaxy is not exactly edge-on. The way  in which spiral arms affect the rotation motion of the gas will be discussed in Section 5.

\begin{figure}
\includegraphics[width=\columnwidth]{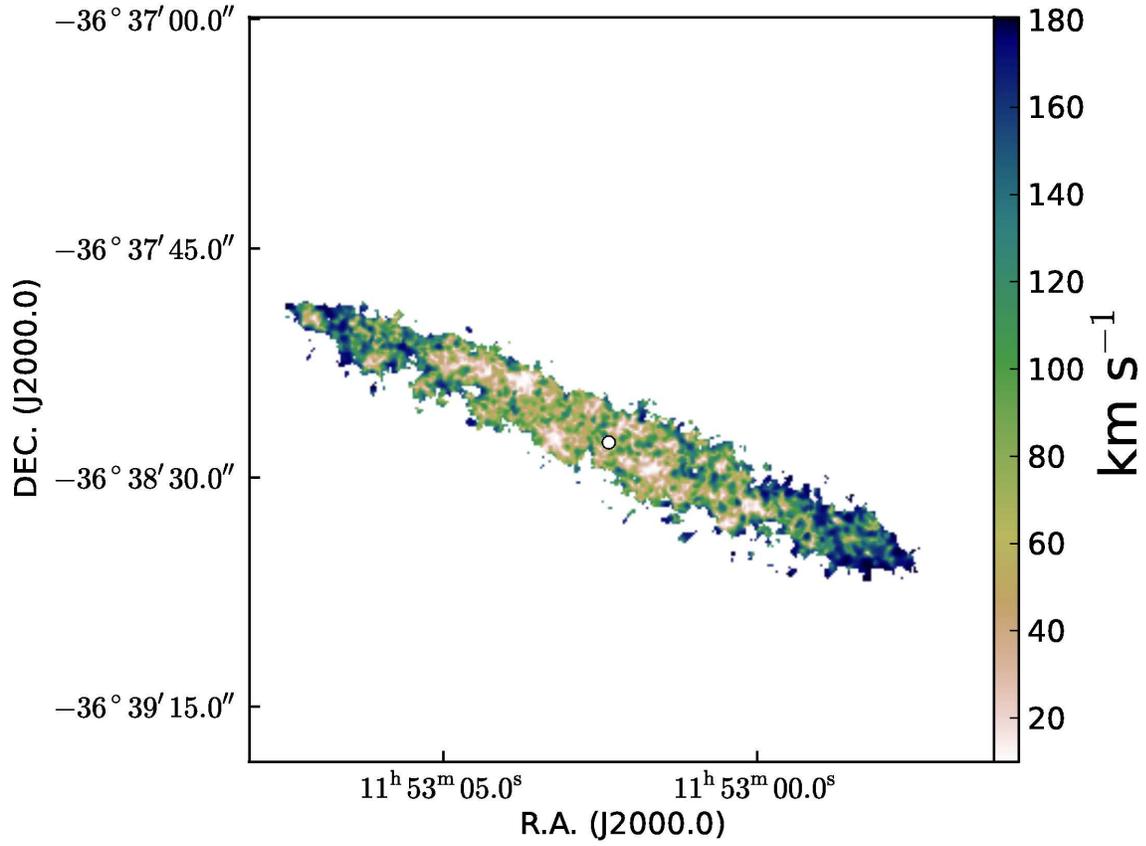}
\figcaption{\label{fig. 6} The second moment map (velocity dispersions) as produced by the task MOMENT. Note the two low velocity dispersion
 bands parallel to the major axis. The white dot stands for the photometric center.}
\end{figure}

\subsection{Position-Velocity Diagrams}

We have built Position-Velocity Diagrams (PVDs) from our data cube or FP $\lambda$-maps ( a cube of dimensions: x, z, $\lambda$). Before extracting the PVDs we have subtracted the continuum to the data cube and rotated it in a way that the major axis (with PA = 67$\arcdeg$) coincides with the X axis (Figure 7).

The PVDs appeared truncated and some of the information was displayed with a sudden discontinuous change whereas they appeared as a part of the same continuous structure. These sudden changes are due to emission with a velocity range larger than the free spectral range (FSR) of the FP interferometer that truncated the structures at that particular velocity range, the FSR. Taking into account that the velocity values measured with a FP interferometer are not absolute but only congruent modulii the FSR and since the FP interferometer used for these observations has a FSR of only 493 km~s$^{-1}$, we produced a larger data cube (of 72 channels) in such a way that the PVDs were not truncated but constructedin reproducing the continuous structures. This was achieved by repeating some of the original velocity channels in the new data cube (of 72 velocity channels) by assigning to these repeated channels the immediate inferior or superior interference order to the 0th order (thus, having a velocity given by the original value from the Doppler effect, minus or plus 493 km~s$^{-1}$, respectively). Thus, the 72 channels-cube has an `extended' FSR equivalent to 887 km~s$^{-1}$. 

 From Fig. 7, we noticed a possible contamination of the [NII] line at $\lambda$ 6548 \AA, (an incomplete, very faint, separate PVD, following the shape of the bright H$\alpha$ PVD, located  about 1/3 the FSR or the separation between the bright PVDs). We checked this possibility considering the following: to isolate the spectral line of interest (in this case H$\alpha$, i.e., with rest wavelength of 6563 \AA) it is necessary to use an order blocking filter. We have used an interference filter whose peak transmission is centered at $\lambda$6630 \AA~ with FWHM of 20 \AA~ (after correcting that value due to the fact that the filters are placed in the convergent F/8 beam of the telescope). The above filter transmission is well suited to isolate the redshifted H$\alpha$ line at the systemic velocity of ESO~379-006  (6627~\AA) because, the nearest emission line to H$\alpha$ is the [NII] ($\lambda$~6548 \AA) line which is about 15~\AA~ to the blue and thus, it should not be visible, in principle, if the filter has a FWHM of 20~\AA~ because usually this line is much fainter than H$\alpha$, and the filter transmission at that wavelength should be quite low. However, filter aging shifts the peak transmission to the blue and also, as discussed in the Introduction, the DIG has [NII]/H$\alpha$ line-ratios larger than those of classical HII regions. Thus, even for the [NII] ($\lambda$~6548 \AA) line which is 1/3 the intensity of the [NII] ($\lambda$ 6583 \AA) line, it could be possible that the [NII] ($\lambda$ 6548 \AA) line contaminates the H$\alpha$ emission if the [NII]/H$\alpha$ line-ratio is larger than usual or if the filter peak transmission has been shifted to the blue several \AA~ due to aging of the filter (or other temporary blue-shifting due to temperature changes or convergent beam effects that are negligible in this case). One of the two possibilities seems to be happening in our midplane PVD at very faint levels (less and around 3$\sigma$; see Figure 7) where there are some hints of a very faint, incomplete PVD {\it separated} in about 180 km~s$^{-1}$ to the blue, following the shape of the bright H$\alpha$ PVD. There is no contamination from the other [NII]  line at $\lambda$ 6583 \AA~ (which is 3 times brighter than the 6548 \AA~ line) probably because of the shifting to the blue of the filter transmission peak and because this other line differs in 20 \AA~ from the  H$\alpha$ line and thus, it does not enter in the filter band.

 In order to subtract the contamination caused by the possible [NII] line,  we carried out a velocity decomposition of the velocity profiles of the original data cube (of 40 channels) in two Gaussian components corresponding to the H$\alpha$ and [NII] ($\lambda$ 6548 \AA) lines. From this profile decomposition,  we have built two data cubes, one for each line. We thus subtracted the data cube of the [NII] line to the original velocity cube. Once we have obtained a corrected data cube we followed again the procedure described before (increasing the velocity channels to 72) in order to enlarge the FSR of the PVDs.


\begin{figure}
\includegraphics[angle=0,width=\columnwidth]{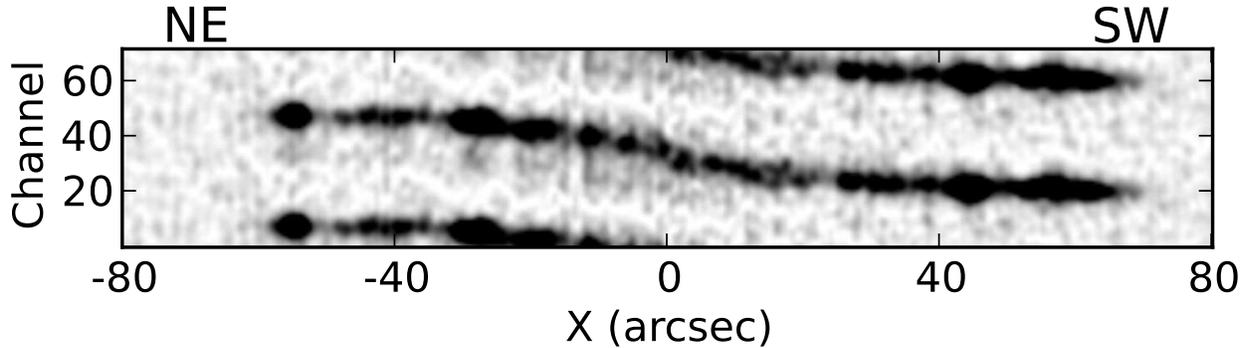}
\figcaption{\label{fig. 7}Image of the PVD along the major axis of  ESO~379-006  extracted from the FP data cube. The central PVD is the complete one, constructed by correcting some features from the free spectral range of the FP interferometer used as described in the text. The incomplete PVDs appearing at the top right and bottom left correspond to artifacts of the reduction software which produces spectra with a wrap-around effect so that emission extending off one end of the spectrum is shown at the other end. Note the detection of diffuse gas (``beard'') at lower rotation velocities (coherent features departing from the bright emission towards lower velocities), as well as the contamination of the [NII] line at $\lambda$ 6548 \AA, (an incomplete, very faint, separate PVD, following the shape of the bright H$\alpha$ PVD, located  about 1/3 the free spectral range or the separation between the bright PVDs) as discussed in the text. The axes are marked in Figure 8.}
\end{figure}

\clearpage



\begin{center}
\includegraphics[scale=0.5]{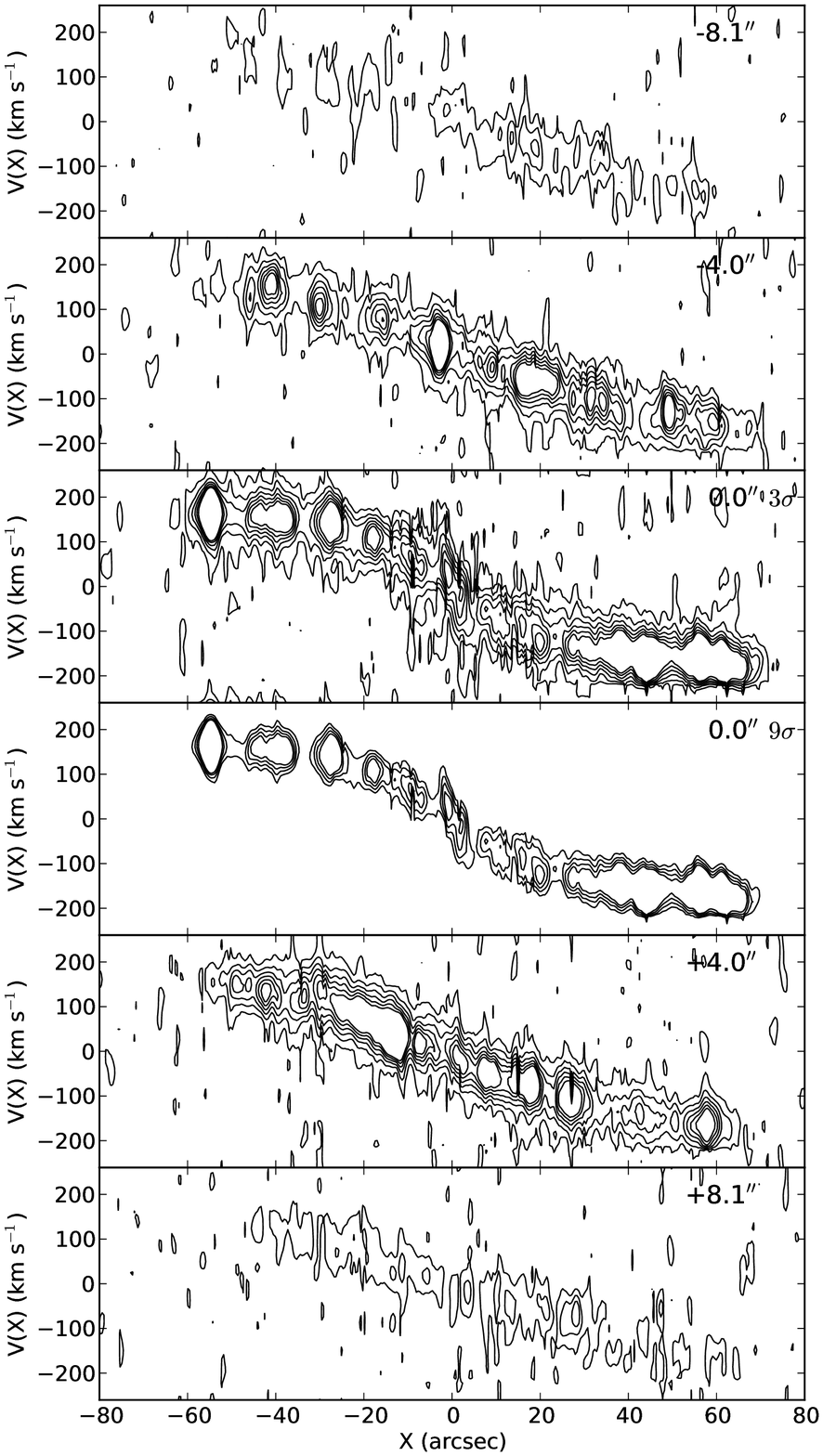}

\figcaption{\label{fig8}PVDs of  ESO~379-006 along several slices parallel to and including the Major Axis of the galaxy (whose Z-value is marked at the top right), visualized as equal intensity contours. These were obtained from the data cube free of [NII] contamination. The X axis is along or parallel to the major axis while the Z axis is perpendicular to it.  The Z
 = 0$\arcsec$ panels show the midplane PVD at 9 $\sigma$ (lower, middle panel; depicting the main disk rotation) and 3 $\sigma$ above the background level (upper, middle panel; showing low velocity extensions interpreted as the ``beard'' e-DIG emission). The other panels are plotted at 3 $\sigma$ above the background level with steps of 3 $\sigma$, without showing the high emission levels. Negative X values correspond to the NE and positive X values to the SW halves of the galaxy. }
\end{center}


\subsubsection{PVDs along or parallel to the major axis}

Figure 8 assembles the contour plots of the PVDs extracted along several Z values of ESO~379--006.  The levels shown in the plots are at least 3$\sigma$ above the background unless otherwise is stated. One of the PVDs was taken along Z = 0 and four others along Z = $\pm$ 4$\arcsec$.0 and $\pm$8$\arcsec$.1, respectively (or $\pm$ 763 pc and $\pm$1544 pc, respectively). All the PVDs correspond to cuts  one-pixel wide. Note however that the data have been smoothed with a Gaussian of 3 pixels (or 218 pc). Note that having in mind Figures 2 and 3, the  values of Z considered do not correspond to true heights above or below the disk, but they could give us some insight about the e-DIG kinematics.

Looking for low intensity gas present at the line of nodes, possibly masked by the bulk of the emission of the projected disk, we have displayed the midplane PVD (Z = 0$\arcsec$, in Figure 8) in two ways: one with levels starting at 9$\sigma$ above the background delineating the motion of the bulk of the H$\alpha$ gas in the disk and the other starting at 3$\sigma$ above the background in order to detect the very faint H$\alpha$ emission.

The midplane PVD  shows gas with velocities spanning from +2641 km~s$^{-1}$  to +3206 km~s$^{-1}$  and a systemic velocity of +2944 km~s$^{-1}$. This later quantity has been obtained by superposing the PVD part of the receding quadrant to the PVD part of the approaching quadrant and trying to render them symmetrical by adjusting the heliocentric velocity in order to match both parts. In the case of ESO~379--006, the approaching and receding quadrants give quite similar and symmetrical PVDs. In obtaining this symmetric PVD, we have also shifted the center of rotation relative to the photometric center of the galaxy (the maximum in the continuum image) in 1$\arcsec$.8 to the NE as listed in Table 2 (offset of kinematical center from photometric center).  Here and the rest of the figures the origin of coordinates is at the kinematical center.

\includegraphics[width=\columnwidth]{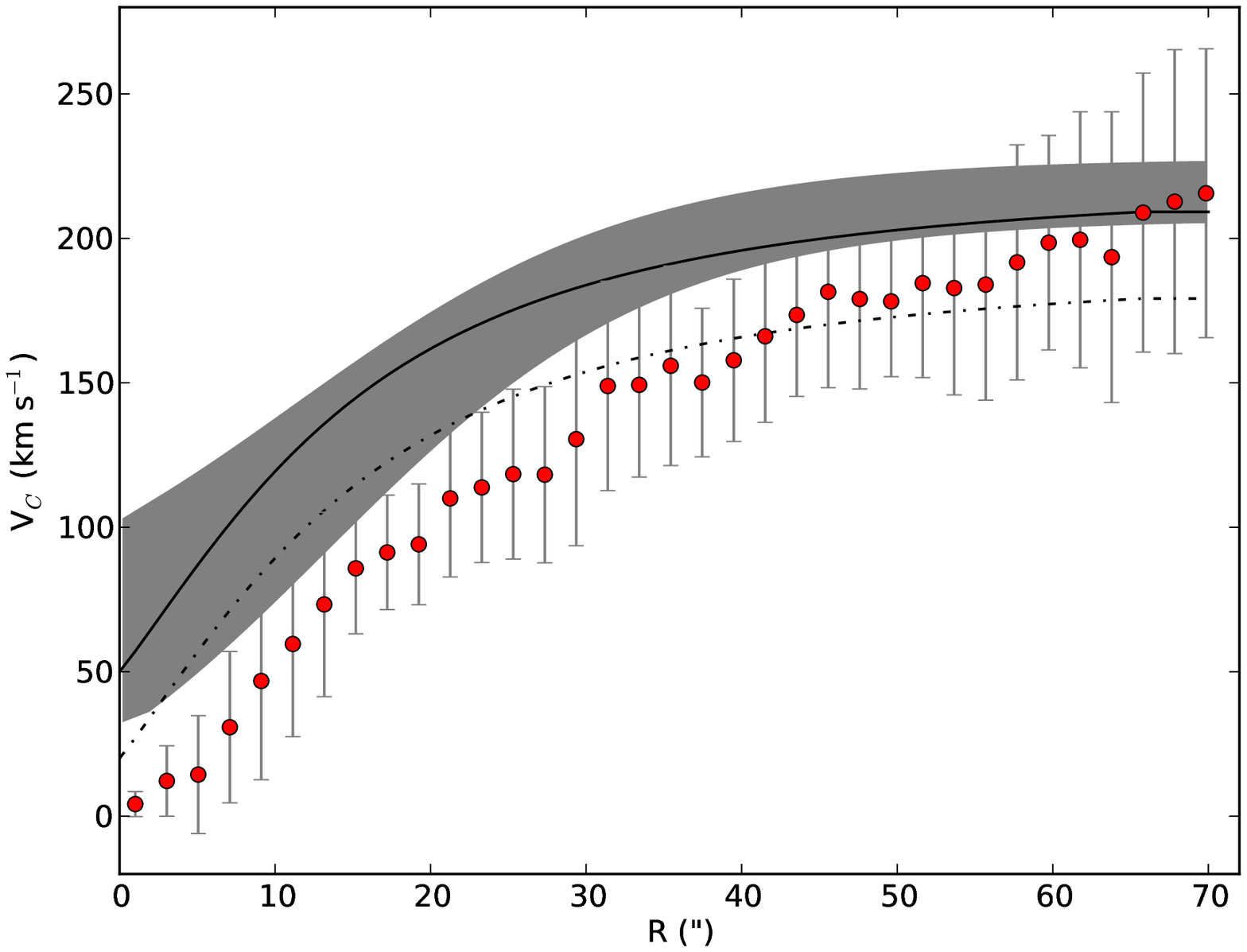}
\figcaption{\label{fig. 9} Rotation curves of ESO~379--006. Filled circles are the barycentric RC with the corresponding error bars
 representing pixel value dispersions within 20$\arcdeg$ azimuthal sectors  2$\arcsec$ long (equivalent to 363 pc) along the major axis.
 The solid curve is the ET RC (i.e., the  $V_{\mathrm{C}}$ of Eq. 2, without correction for $\overline{\sigma}_{\mathrm{r}}$); the shadowed area is bounded by the fits to the RCs at
 $I_{\mathrm{env}}\ \pm\ 3\ I_{\mathrm{min}}$.
 The dash-dotted curve is the $V_{\mathrm{C}}$ given by Eq. 2 (i.e., corrected for $\overline{\sigma}_{\mathrm{r}}$).}

\subsubsubsection{The Rotation Curves}

The classical method for determining the rotation curve from a velocity profile is called the intensity peak method. It consists of fitting the profile using a theoretical function like a Gaussian or alternatively, in computing directly the barycenter of the line, the goal being in both cases to measure the wavelength of the intensity peak which provides the mean velocity of the line. In case of edge-on galaxies, the classical method does not work because the shape of the profile is strongly
affected by line-of-sight (LOS) effects that tend to underestimate the true rotation velocities and consequently, some corrections to the estimated velocities should be done in order to really map the rotation and to infer the gravitational potential. These effects arise because, at each galactocentric distance along the major axis of an edge-on spiral, the LOS includes several different parcels of gas with different projected velocities along the LOS, spanning continuously (if the disk is uniform) from the full rotation (for gas at the line of nodes) to the systemic velocity (for gas orbiting perpendicular to the line of sight). An sketch illustrating this effect can be found in Garc\'ia-Ruiz el al. (2002, see their Figure 2).
Consequently, to obtain the rotation curve (RC) of an edge-on galaxy is not as straightforward as in the case of less inclined galaxies.

Two main methods have been proposed to correct for LOS effects in edge-on galaxies: a) the Envelope-Tracing, ET,  Method (developed by Sancisi \& Allen, 1979, and described in van der Kruit, 1989 and in Sofue \& Rubin, 2001), and b) the Iteration Method (Takamiya \& Sofue 2002). Both of them start by an analysis of the position-velocity diagrams (PVDs) along, parallel to or perpendicular to the major axis of the galaxy. Since, according to Section 3, this galaxy is not exactly edge-on, the PVDs we have built  serve to examine to what extent possible  LOS effects are present and how to take them into account.

The midplane PVD  (middle panels of Figure~8) was used to apply both, the Intensity Peak (or ``barycentric'') Method and the Envelope-Tracing Method to this galaxy in a similar way as Sancisi \& Allen (1979) did for obtaining the midplane rotation curve of NGC~891. We derive the barycentric RC from the first moment map taking an angular sector of $\pm$ 20$\arcdeg$ along the major axis. We should remind that the first moment map was constructed by taking as velocities the intensity peaks of the velocity profiles determined pixel per pixel. 

On the other hand, we apply the ET Method as formulated in Sofue \& Rubin (2001) and used for HI gas by Heald et al. (2006b). Namely, the intensity of the envelope
 on the line profile is chosen according to:
\begin{equation}
I_{\mathrm{env}}=\sqrt{(\eta I_{\mathrm{max}})^2+I_{\mathrm{min}}^2},
\end{equation}
where $\eta$ is a dimensionless constant with usual values 0.2--0.5, $I_{\mathrm{max}}$ is the maximum intensity in the line profile, and $I_{\mathrm{min}}$
 is the minimum value of the contour taken to be $3\sigma$ of rms noise of the H$\alpha$ PVD diagram.

In this way, the rotation velocity can be determined as follows:
\begin{equation}
\label{eq1}
V_{\mathrm{C}}=(v_{\mathrm{env}}-v_{\mathrm{sys}})/\sin{i} - \overline{\sigma}_{\mathrm{LOS}} ,
\end{equation}
where $v_{\mathrm{env}}$ is the velocity of the envelope at $I=I_{\mathrm{env}}$, $v_{\mathrm{sys}}$ is the systemic velocity,
 $\overline{\sigma}_{\mathrm{LOS}} = ({\sigma^{2}_{\mathrm{inst}}+\sigma^{2}_{\mathrm{gas}}})^{1/2}$ is the average velocity dispersion due to
 the instrumental velocity resolution, $\sigma_{\mathrm{inst}}$, and the gas velocity dispersion, $\sigma_{\mathrm{gas}}$.
 We chose the following values for the parameters: $\eta=0.3$ sufficiently low in order to take into account the high velocity wings of the disk profiles (fainter than the bulk disk emission) and $\overline{\sigma}_{\mathrm{LOS}} = 30$ km~s$^{-1}$ as the expected value for an instrumental resolution of 25 km~s$^{-1}$ and a thermal width of 12  km~s$^{-1}$.   Figure 9 displays both the barycentric and the ET rotation curves of ESO~379--006 as obtained from the PVD.
 These curves will be used in the galaxy modeling discussed in Section 4.3.

While the barycentric RC is less steep at R $\le$ 40$\arcsec$ (equivalent to 7.3 kpc)  than the ET RC, both curves reach similar maximum rotation velocities at larger radii (of about 200 km~s$^{-1}$), larger than the value reported by Mathewson et al. (1992) and Persic \& Salucci (1995; see Table 2). When $\sigma_{LOS}$ is taken into account in the ET method, the resulting RC is traced by the dashed line in Figure 9. In conclusion, LOS effects are important for R $\le$ 40$\arcsec$; for larger galactocentric distances both RC are similar.

\subsubsubsection{The low intensity gas}

The PVD along the major axis (MA-PVD) shows extensions of faint gas (usually
 detected at 3 $\sigma$ levels)
having  lower motions than the bulk rotation of the disk traced by the bright
 emitting gas. Figure 7 and the upper, middle panel of Figure 8 show this.
The zero velocity corresponds to the systemic velocity and the negative X values correspond to the receding part whereas the positive X values correspond to the approaching part of the disk. As usual, for reference, the PVD can be divided in four quadrants: 1, top-left, 2, top-right, 3, bottom-left and 4, bottom-right.

In this particular case, in Figure 7 and Figure 8 (third panel from top to bottom),
 we do not refer to the ``normal''
width of the PVD, followed over all the PVD as a kind of velocity dispersion,
  but to very localized extensions (or peaks) sorting out in an irregular way
 from the main body PVD. In that context we note the large extensions near the
 center (from --10$\arcsec$ to +10$\arcsec$), the low velocity peaks in the receding side
 of the PVD between --20$\arcsec$ to --50$\arcsec$ and the low velocity extensions in the
 approaching side between +10$\arcsec$ to +70$\arcsec$. In order to be sure of the reality of those features, we will not consider in this
 discussion the extensions seen in the receding side because it is
 possible that the subtraction of the [NII] line could not be as proper as it 
is required and some residuals could still contaminate the contours. Thus, we
 will concentrate on the extensions of faint gas visible on the approaching side
 (corresponding to the SW side of the galaxy) where there is no possible
 contamination of the [NII] line (that line would appear in the lower side of
 the PVD and the extensions are seen in the upper side). For the approaching
 side (SW) of the PVD, the extensions appear 
only with lower velocities than the bulk rotation and it is noticeable that 
no extension at velocities higher than the bulk rotation is detected. In
 particular, we notice the extensions between +10$\arcsec$ to +20$\arcsec$ and the
 extensions between +50$\arcsec$ to +70$\arcsec$ from the galactic center, seen in 
the MA-PVD at 3$\sigma$ (Figure 8, third panel).

For an almost edge-on galaxy, the existence of low velocity extensions has 
been generally attributed to beam smearing and/or LOS  
projection effects. 

As mentioned before, LOS projection effects (and also beam smearing) consist in that the velocity profile at a given point 
of the galaxy, instead of being the profile of a single parcel of gas at a 
given galactocentric radius, is the result of the integration  through the 
LOS of several parcels of gas of the galaxy disk having each parcel different
 LOS projected velocities. This integration over several velocities is due
 either to a large beam size or to geometry, in the case of edge-on galaxies.
 This later possibility arises because, even if the angular resolution is good,
 the line of sight crosses through all the disk gas at a certain projected
 distance from the center. The resulting velocity profile for a given distance
 to the center, X, is the integration along the LOS, weighted by the intensity,
 of the projected LOS velocities of different parcels of gas with different
 densities across a large range of galactocentric radius. The resulting
 velocity profile is not Gaussian and the velocities vary from the actual
 rotation velocity (from gas parcels at the line of nodes) to values approaching
 zero (thus, approaching the systemic velocity). In the ideal case where the brightness
 distribution of the disk is uniform, axisymmetrical and decaying exponentially
 with radius, the true rotation velocity at a given X would correspond to the
 highest velocity gas detected in the velocity profile. This effect is
 particularly important in the central parts of the rotation curve where LOS
 effects imply the integration across parcels of gas with more different
 projected velocities than in the external parts of the rotation curve. That
 explains why in edge-on galaxies the ``barycentric'' RC does not represent 
the true rotation in the inner parts of the galaxy. The treatment of the
 velocity profiles in order to ``rescue'' the high velocities, that really
 measure the actual rotation, constitutes the basis of the ET
 Method for edge-on galaxies as described before and applied in 
Section 4.2.1. In that later subsection we have demonstrated that LOS
 effects are important for ESO~379--006 because the
``barycentric'' and the ET rotation curves differ (see Figure 9).
As a bi-product, low velocity extensions approaching the systemic velocity 
would be observed in the major axis PVD, giving it a morphology that resembles 
a  
``butterfly wing''. We called that morphology that way because, those extensions appear as a $\it continuous$
 velocity decrease mostly concentrated towards the central regions as
 the  velocities vary continuously depending on the  projection along the LOS
 and, at the central regions the LOS crosses more parcels than near the edges.
 We reproduce, in the next sections, the morphology of MA-PVDs with this LOS effect in the models
 we have run.

It is that interpretation that is currently used to explain the faint, low
 velocity extensions in MA-PVDs of nearly edge-on galaxies. However, low
 velocity, faint extensions have  been detected in less inclined galaxies such
 as NGC~2403  ( with $\it i$ $\simeq$ 60$\arcdeg$) in its MA-PVD obtained from the 
HI gas, and other spirals (Fraternali et al. 2001, Heald et al. 2011,
 respectively). In the HI PVD along the major axis of NGC~2403, low velocity
 extensions (and even forbidden velocities) are detected in such a conspicuous
 way that those authors called ``the beard''. For the relatively low inclination
 of that galaxy, no LOS effects are expected (unless the beam of the HI
 observations was of very low angular resolution, which is not the case).
 Besides, the low velocity gas appears as localized extensions, not present
 everywhere. Fraternali et al. (2001) explain those low velocity extensions as
 corresponding to low density gas above or below the disk, having rotation
 velocities lower than the disk rotation, or even presenting a gradient in
 rotation velocity  with vertical height (this situation is called ``lagging 
in rotation''). Thus, if the explanation of ``the beard'' as vertical gas with
 lagging rotation, seen projected on the major axis  is true, that beard should
 also appear for more inclined galaxies. Indeed, 
in more inclined galaxies (except the case of $\it i = 90 \arcdeg$), it should appear a combination of both effects in their MA-PVD:  the
 ``butterfly wing'' (due to LOS projection effects) and ``the beard'' (due 
to lagging vertical gas) as low velocity extensions. 
 However, till now, 
nobody has attributed a vertical origin to some of the low velocity extensions
 detected in more inclined galaxies. 

The MA-PVD of ESO~379--006 depicted in Figures 7 and 8 shows low velocity
 extensions that resemble very much the morphology of the MA-PVD of NGC~2403
 though the first was obtained for the ionized gas and the second, in HI. Since
 ESO~379--006 is viewed more inclined than NGC~2403 (while it is not completely
 edge-on), it is worth to analyze how the detection of ``the beard'' on the
 MA-PVD depends on different factors.
Following the
 interpretation by Fraternali et al. (2001), it would depend first of all, 
on the existence of extraplanar gas with an actual difference in rotation
 relative to the midplane rotation (lagging) or at least with a marked 
gradient in the rotation velocity with vertical height (Z). Assuming that 
that requirement is fulfilled, it would also depend on geometrical factors
 involving the galaxy inclination and the maximum vertical height of the
 extraplanar gas (Z$_{MAX}$)
(see Sect. 3 and Table 2).  We refer
 again to Figure 2 where we display the  H$\alpha$ image with 
overlayed contours of equal surface brightness of bright HII regions in the 
disk of the galaxy as well as a geometrical 
sketch considering the e-DIG distributed in a cylinder extending up and down 
the galactic plane with Z$_{MAX}$ equal to the projected plane. There is no minimal inclination required for detecting ``the beard'' whereas,
 if 
$\it i$ is exactly 90$\arcdeg$, no beard would be detected in the MA-PVD
 because the gas in the vertical direction is seen outside the major axis.

In the particular case of ESO~379--006, since $\it i$ = 82$\arcdeg$.5, it is possible to have both effects present: LOS projection and extraplanar DIG lagging revealed by different shapes in the low velocity extensions of the MA-PVD. In the case of the prototypical edge-on galaxy NGC~891,
Swaters et al. (1997)
 determine an inclination for NGC~891 of at least 88$\arcdeg$.6. These authors also show a MA-PVD in HI looking more like the ``butterfly wing''
 appearance, more characteristic of LOS projection effects. Besides, that work
 reveals the photometric asymmetry of that galaxy. Thus, 
the absence of beard detection in the MA-PVD, confirms that NGC~891
is more edge-on than ESO~379--006.

In conclusion, the MA-PVD of ESO~379--006 shows faint extensions approaching 
the systemic velocity of the galaxy. As mentioned before,  that kind of
 extensions are usually attributed to LOS projection effects in nearly 
edge-on galaxies. However, the fact that this galaxy 1) is not exactly edge-on,
 2) that it has been demonstrated by other means that localized e-DIG is
 detected up to vertical heights that can be seen projected on its major axis
 and 3) that, in what concerns the low velocity extensions, the MA-PVD 
resembles more the beard-type and not so the butterfly wing-type because 
the faint extensions are localized (as the e-DIG in this galaxy is) makes us 
to propose that some part of these extensions are due to e-DIG lagging in
 rotation relative to the disk although we cannot fully discard the possibility that those extensions are due solely to LOS projection effects of a highly
 inhomogeneous disk.

\includegraphics[width=\textwidth]{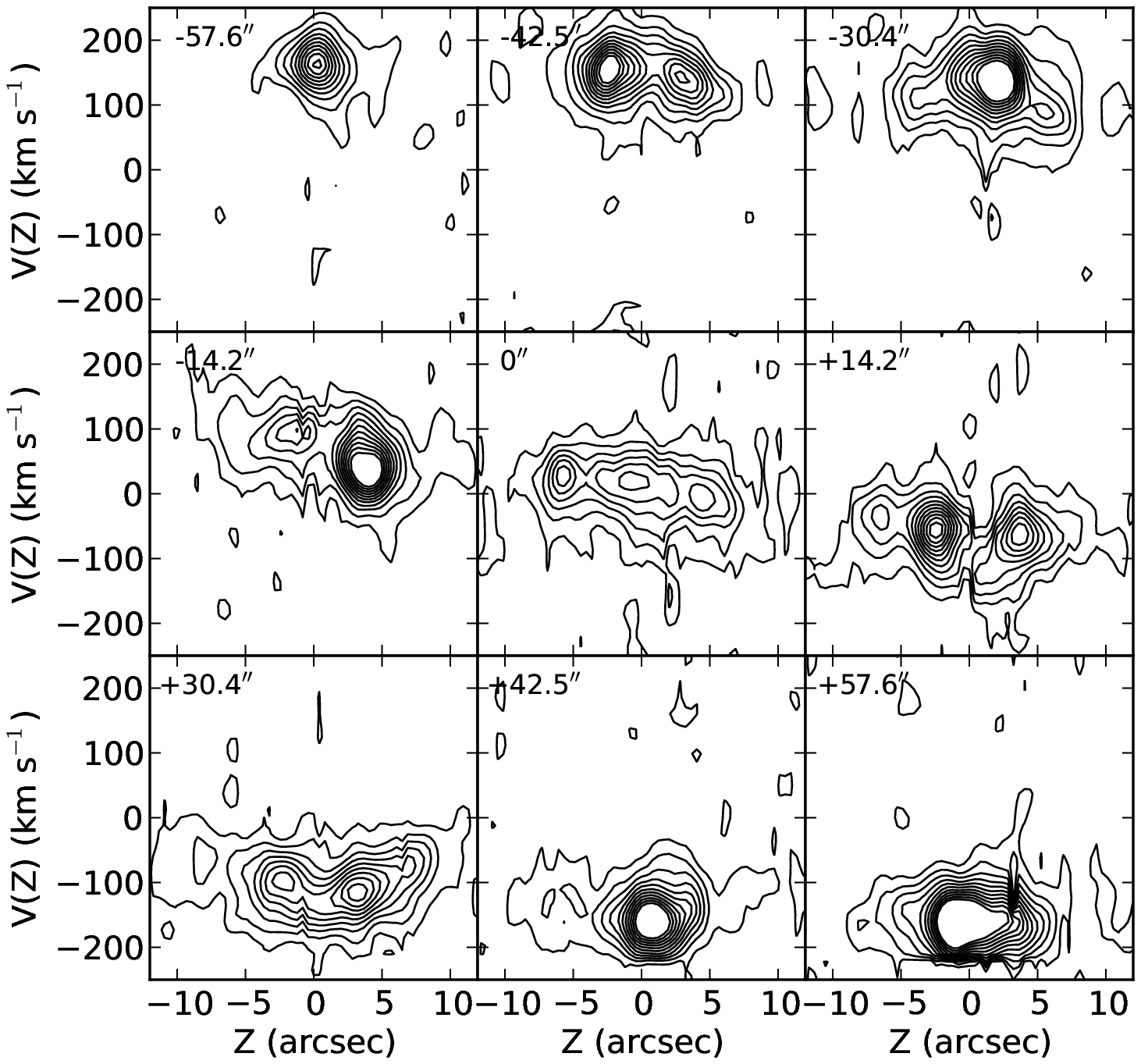}
\figcaption{\label{fig. 10}Observed PVDs parallel to the geometrical minor axis. Contour levels start above 3$\sigma$ of the background level, with steps of 3$\sigma$. The different PVDs were constructed for different X values (marked at the top-left). The sense of the Z-axis is such that the negative values correspond to the nearest side of the galaxy (the Southern side).
 Note the difference in velocities between the two intensity maxima corresponding to the panels where X $\le \pm42.5 \arcsec$ (equivalent to  $\pm7.7$ kpc). }

\clearpage

\clearpage

\subsubsection{PVDs parallel to the minor axis}

PVDs were also obtained along and parallel to the minor axis. These were constructed under the assumption that the minor axis of ESO~379-006 is perpendicular to the geometrical major axis. Figure 10  shows the resulting PVDs.
The PVDs were extracted from the data cube corrected for the [NII] line contamination (see Section 4.1) with cuts one pixel wide.

As one can see from these figures, the high intensity emission forms a 
heart-shaped profile as it was found in models published by Fraternali \& Binney (2006; see their Figure~15). We can also see that the  shape inverses in the top-down direction and when one passes to the cut at X = 0$\arcsec$ we corroborate that this cut  corresponds to the minor axis of the galaxy because it is the most symmetrical one. An inspection of the central PVDs (--42.5$\arcsec \leq$ X $\leq +42.5\arcsec$; i.e., within 7.7 kpc ) reveals that, for those PVDs, it is detected a velocity asymmetry of the bright emission consisting in a velocity difference of about +25 km s$^{-1}$ between the SE (low) and the NW (high) sides  of the disk in the heart-shaped profile.
We have inspected PVDs extracted at several other positions in order to check the general trends of the velocity asymmetry and we have found that the velocity difference in the PVDs remains almost constant regardless  the position in a region of about 100 pixels (equivalent to 40$\arcsec$.5 or 7.4 kpc) around the center.
In view of this first order deviations from circular motions found for the bright emission, we do not search for any further deviations of the faint emitting gas (such as the gas producing ``the beard'' in the PVD along the major axis) because these second order deviations would be affected by the  non-circular motions of the bright-emitting gas.

\subsection{Disk modeling with GALMOD}

\subsubsection{ Galaxy models}

 As mentioned in Section 4.2, in order to study in detail the kinematic features of this galaxy, 
we extracted several PVDs parallel and along the major axis (Figures 7 and 8 )
 and parallel and along
the minor axis (Figure 10). Due to its  inclination, we also explored possible
 LOS projection effects by obtaining both Intensity peak (Baricentric) and
 Envelope-Tracing (ET) Rotation Curves of this galaxy. From these curves we have
 found that LOS projection effects are important for R$\le 40 \arcsec$.
Since LOS effects are present, the way of obtaining the main parameters of
 this galaxy is not as straightforward as in less inclined galaxies. 
We used the routine GALMOD  of the GIPSY package in order to determine realistic
 parameters of this galaxy, because this task allows us to built a model and
 take into account projection effects.

The task GALMOD is useful in confronting kinematical data produced from
 observations (such as velocity maps, radial velocity fields, PVDs, rotation
 curves, etc.) with models of a gaseous disk assumed as a series of concentric
 tilted rings where both 
 inclination and position angle of the different rings may vary or remain
 constant (``tilted rings''). The task was developped to confront HI kinematical observations with
the models but several authors have applied it also to treat ionized gas
 kinematics as well.
The gas is assumed to be in circular orbits on these rings.
Since the tilted ring disk models have as parameters: a circular velocity, a
 velocity dispersion, a column density, a scaleheight perpendicular to the plane
 of the ring, an inclination and a position angle, we can vary every one of
 these parameters and study their behaviour in the synthetic generated velocity
 maps, radial velocity fields and PVDs.
The goal is to control different parameters of the galaxy model in order to fit
 the synthetic cubes with the observed data cubes.
We have convolved our modeling results in such a way that their linear
 resolution matches with the resolution of the observed data (3 pixels,
 corresponding to 1.$\arcsec$2 or 218 pc). Also, as a test, we have smoothed the
 modeled cubes to a resolution of 5 pixels (equivalent to 2$\arcsec$ or 363 pc) 
and we do not see any important change in our results nor in the confrontation
 with observations.

In order to accomplish that we defined a model of the disk. This model disk 
was
constructed as a series of concentric annulii or rings (we prefer to call them
 annulii in order to exclude possible confusion with the ring of HII regions
 found in ESO~379--006, see Section 3) where both the PA and 
inclination were assumed constant with radius. In this context, the planes of the annulii coincide with the plane of 
the disk.
 To each annulus we asign a circular velocity $V_C$ at the corresponding radius
 of the annulus  obtained from our RCs (ET and barycentric) and a systemic
 velocity
 $V_{sys}=2944$ km s$^{-1}$, obtained after trial and error considering other
 values and taking the value that makes both sides of the RC symmetrical. We
 initially assigned the intensity-peak or
 barycentric RC (filled circles in Figure 9) however, 
we find that the models agree better with the observed data if we use the ET RC
 instead of the Barycentric RC, as expected.
We also asigned as a start
 a constant LOS  
velocity dispersion,  $\sigma_{LOS}$ =30 km s$^{-1}$ -- as expected according 
to the discussion on the quantities appearing in Equation (2).

We have also used as a start a gas radial distribution consisting of an
 exponentially decaying disk and later on,
 in an attempt 
of considering more realistic radial distributions (in brightness not in mass)
 we have added to the exponential gas disk an axisymmetric ring. We explored 
the models described in what
 follows varying also the ring parameters, however, the change of those
 parameters has no apparent effect on the velocity field.

The vertical scaleheight values were varied in such a way that  the model disk
 with such parameters when projected, roughly reproduces the observed image.

Finally, as it will be discussed below, we find the necessity for including
 a radial motion in our models.
 For this purpose we have modified the GALMOD routine to include an additional
 velocity component, the radial velocity $V_{R}$, such that the LOS velocity for
 each annulus is given by:
\begin{equation}
\label{eq2}
V_{\mathrm{LOS}}=V_{\mathrm{sys}}+V_{\varphi} \cos{\varphi} \sin{i}+V_{\mathrm{R}} \sin\varphi \sin i ,
\end{equation}
 where $\varphi$ and $i$ are the azimuthal and inclination angles, respectively,  and $V_{\varphi}$ is the azimuthal velocity that comprises the circular velocity
 and LOS velocity dispersion, $\sigma_{\mathrm{LOS}}$ as given by Equation 2. 
On the other hand, we do not include any $V_{\mathrm{Z}}$ component in our
 models. 

We have thus run a grid of GALMOD models with different parameters in order
to confront the synthetic results with the observed ones and to try to have some
 control on what each specific parameter change in the synthetic results.
The grid consists in varying the following parameters: 1) circular velocity 
(ET or barycentric), 2) velocity dispersion (30, 40 km/s), 3) radial
 distribution of scale radius of 200$\arcsec$ (measured from the observed
 images) but considering an exponential disk or an exponential disk plus a ring (of different radii, 24$\arcsec$ and 60$\arcsec$ and amplitudes of 0.4 and 8.6),
 4) vertical scaleheight (0. i.e. a disk with no width,  1$\arcsec$.5,
 2$\arcsec$, 3$\arcsec$, 4$\arcsec$ and 5$\arcsec$), inclination of the disk 
(80$\arcdeg$, 82$\arcdeg$.5, 85$\arcdeg$, 88$\arcdeg$) and several PA's and
 centers but we have quickly converged to the values of PA = 67$\arcdeg$ and the
 center displaced from the photometric center as listed in Table 2. We have also
 used a systemic velocity $V_{sys}=2944$ km s$^{-1}$ while we have explored
 other values.

\subsubsection{ Confrontation with observations}

There are some important features in the observed data that we took into account
 in selecting the more suitable synthetic GALMOD model, the more important are:

-- The bifurcation noticed in the central velocity maps (see Fig. 4).

-- The twisting of the isovelocity contours seen in the SW half of the galaxy
 radial velocity field (Fig. 5).

-- The presence of emission in the PVDs parallel to the major axis even for
 vertical heights Z = $\pm$ 4$\arcsec$.0 and 8$\arcsec$.1.

-- The kinematic asymmetry found in the PVDs parallel to the minor axis 
(Fig. 10).

-- The slight inclination of the minor axis PVD  (Fig. 10).

Taking into account all these aspects we obtained from the different models
 synthetic velocity maps, radial velocity fields, PVDs parallel to the main 
axii of the galaxy and study the general trends that changing one or several
 parameters produce on those synthetic maps.

We have noticed that when comparing the modeled major axis PVD 
with that of the observed galaxy,
the assigned velocity dispersions resulted to be small, giving a  narrower PVD
 than the observed one and large residuals. 
The shape of the modeled major axes PVDs is mainly determined by the form of the
 RC 
and magnitude of $\sigma_{LOS}$.
We found that $\sigma_{LOS}\approx$ 35 to 40 km s$^{-1}$  is required to recover
the width of the major axis PVD. Note that this value is larger than the value
 used in RC modeling ($\sigma_{LOS}$ = 30 km s$^{-1}$); this value is in
 relative agreement with the observed velocity dispersion values displayed in
 Fig. 6.

\clearpage
\noindent
\includegraphics[width=\textwidth]{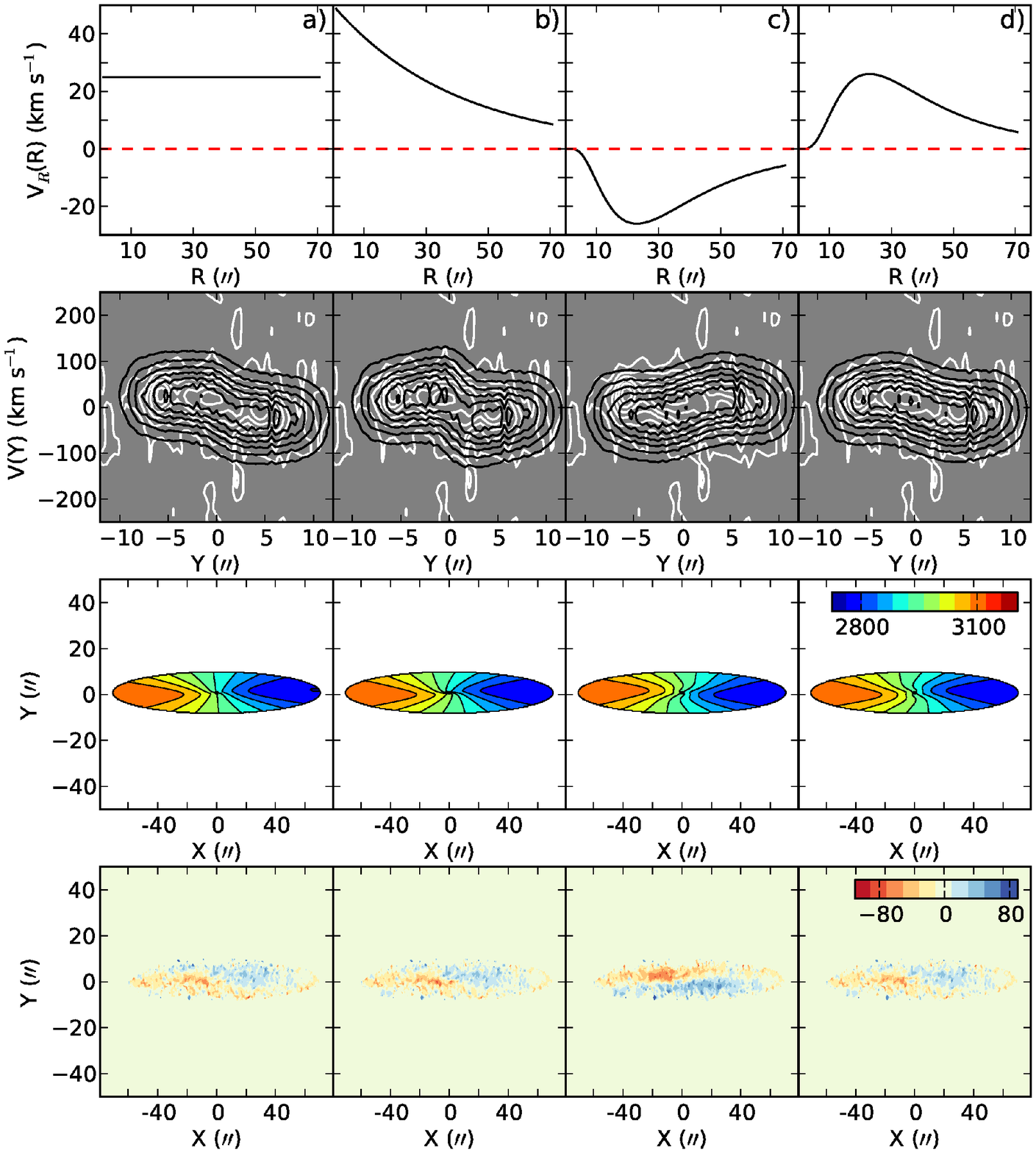}
\figcaption{\label{Fig. 12} The radial motion distribution (top row), a comparison of modeled (black contours) and observed (white contours)
 minor axis PVD (second row), synthetic velocity fields (third row)
 and the residual velocity fields (bottom row) for models  with a constant inflow (a), exponential decaying inflow (b), lognormal outflow (c)
 and the lognormal inflow (d).}
%
\subsubsection{ The necessity of an inflow}

In the case of PVDs parallel to the minor axis, the PVDs extracted from the
 synthetic cube should reproduce the kinematic asymmetry found in the observed
 PVDs (see Figure 10). We have produced synthetic PVDs parallel to the minor
 axis
extracted from the different models  and we have found that in order to
 reproduce
 the kinematic asymmetry seen in the observed PVDs  in Figure 10 we have to
 include a radial motion besides the rotation motion, as discussed before,
 giving rise to the LOS velocity of Equation 3.

We studied several functional forms for the inflow/outflow radial motion
 distribution.
 These are resumed in Figure 11.
 For the models discussed below each row represents (from top to bottom) the
 radial motion as a function of the radius of the disk,
 the minor axis PVD contours overlaid on the observed one,
 the synthetic velocity field generated with VELFI procedure of the GIPSY
 package, and the residuals calculated as the observed minus
 synthetic velocity field.

As first try, the simplest choice is a model with a constant inflow/outflow
 velocity (Figure 11a).
 These models are useful for discriminating between inflow and outflow motion
 since we can compare directly the minor axis PVDs and
 a synthetic radial velocity map in search for the direction of the twisting. 
As we have already determined that the South side of the disk is
 the nearest to the observer (Sect. 3), the direction of radial flow should be
 inwards to the center. 
 Thus, we conclude that in order to reduce the residuals in the modeled velocity
 field and to fit the PVDs parallel to the minor axis it is necessary to add an
 inflow motion. 
 While this model reproduces well the overall shape of the minor axis PVD it
 fails in reproducing the shapes of the PVDs
 parallel to the minor axis at the edges of the disk.

In order to take into account the vanishing of the inflow at large radius we
 adopted the exponentially decaying profile
 $V_{\mathrm{R}}=V_{0}\exp\left(-R/h_{\mathrm{V}}\right)$. A similar  velocity
 profile was measured in a dwarf 
 galaxy NGC 2915 by Elson et al. (2010).
 We have made several models with different values of the maximum velocity,
 $V_{0}$, and
 the scale-radius $h_{\mathrm{V}}$. After iteratively adjusting the values
 of $V_{0}$ and $h_{\mathrm{V}}$ to match the observed velocity field we were
 able to minimize the residuals in the velocity field for $V_{0}=50$
 km s$^{-1}$ and $h_{\mathrm{V}}=40\arcsec$ (equivalent to 7.3 kpc, see Figure
 11b). 

However, radially decreasing $V_{\mathrm{R}}$ cannot
 explain the kinematic features in the center such as the channel map at the
 systemic velocity (see Figure 4) 
 and the inclination of the  PVD along the minor axis. Also, the major axis PVD
 appears smeared
 in the center due to large velocity dispersion. We found that in order to
 reproduce the correct shape
 of the channels or velocity maps close to the systemic velocity, the radial
 motions should be nearly zero at the center. 
 For this reason, we chose to model the inflow as the lognormal distribution:

\begin{equation}
V_{R}=\frac{A}{R w}\exp\left(-\frac{(\ln R/{h_{V}})^2}{2 w^2}\right) .
\end{equation}
We iteratively adjusted the parameters of the model to match the observed data.
 The distribution of the inflow
 velocity that we used as a final model is plotted in Figure 11d
 and has the following parameters: $A=480\arcsec$ km s$^{-1}$, $h_{\mathrm{V}}=35\arcsec$ (or 6.4 kpc), and $w=0.65$.
 Note however, that these values are only tentative ones.

For illustrating purposes only, we also show the outflow model produced by
 changing the sign of $A$ in Figure 11c.
 This model is ruled out in this case because the modeled velocity field does
 not correspond to the observed velocity field,
 that appears twisted counterclockwise in Figure 5. As it can be observed, the
 residuals for the outflow case show an
 excess of the velocity on the opposite side of the disk when compared with the
 panels for the other models (inflow).

Although with the lognormal inflow distribution the minor axis PVD is well
 reproduced, we were unable to eliminate completely the residuals. 
 This may indicate that we do not fully take into account all the non-circular
 motions. We varied the inflow velocity but we have found that the value of this
 quantity is fixed to the value of the kinematic asymmetry in the observed minor
 axis PVDs. Consequently, in order to reproduce them we cannot vary the inflow
 velocity arbitrarily. Even if we cannot fully eliminate the residuals, we
 correctly
 reproduce the direction and the degree of twisting of isovelocity contours. 
 A visual inspection of the residuals on Figure 11 suggests that inflow model d)
 is better than a) and b) because the residuals show less systematics.

 In order to select the best model we further compare several PVDs parallel to
 the main axii and also the individual
 velocity maps of the modeled cube with the observed one. These are displayed 
in Figures 12, 13 and 14, respectively, for the favored lognormal inflow model. The
 comparison of the figures for the velocity maps and PVDs for the major
 and minor axes, reveals an excellent agreement, supporting the validity of our
 assumption of the lognormal inflow
 velocity distribution. Several features seen in these figures should be
emphasized.
The shape and the extension of emission regions and also bifurcations of
 contours in the observed channel
 maps at $6\sigma$ contour levels are satisfactorily reproduced (Figure 14).
See the discussion in Sect. 4.3.4 about the bifurcations.

\begin{center}
\includegraphics[scale=0.55]{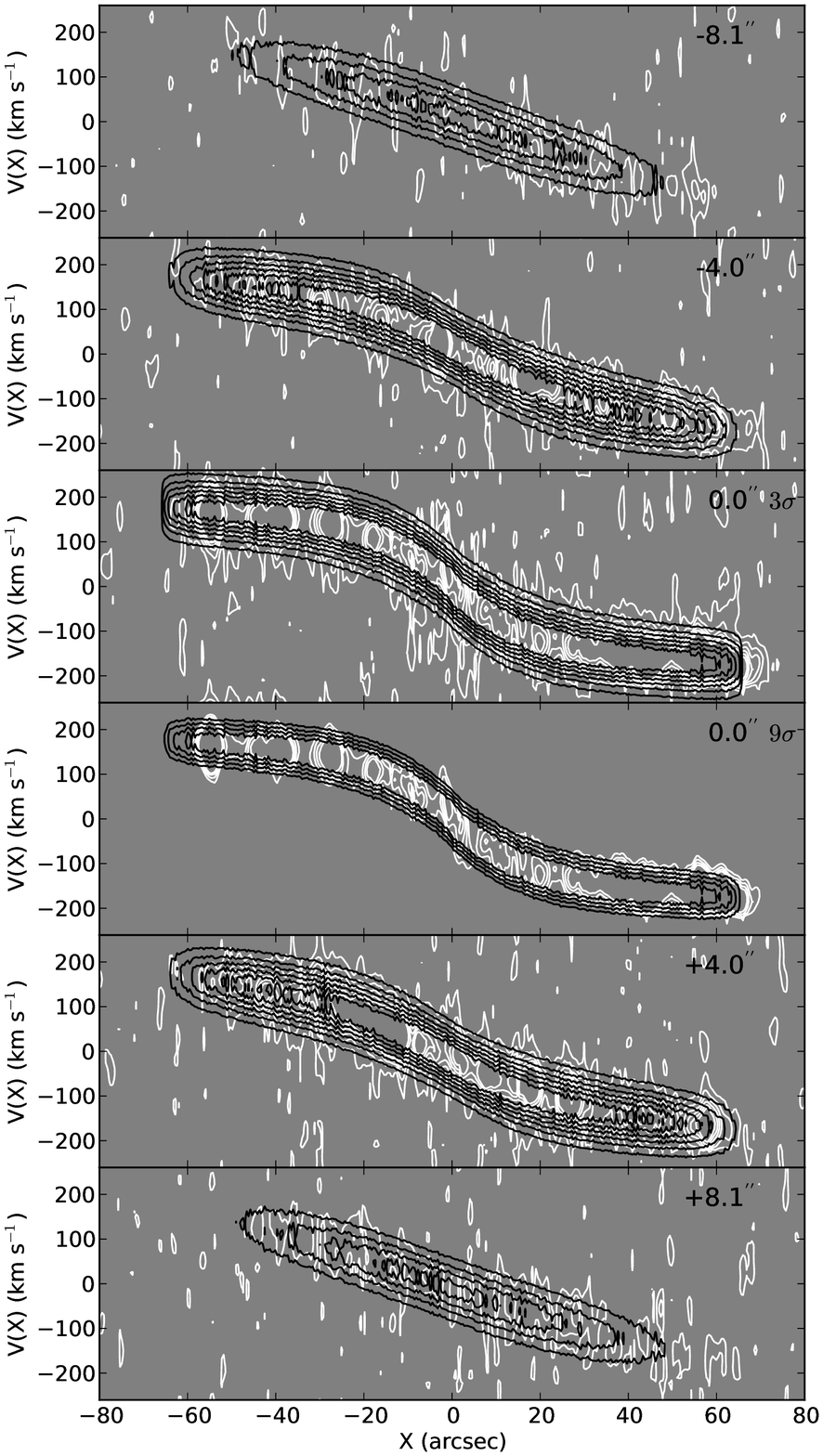}
\figcaption{\label{Fig. 11} Modeled PVDs contours (black) parallel to the major axis including lognormal inflow and an exponential plus ring radial
 distribution (Model R).
 Contour levels are in
3$\sigma$'s of emission intensity normalized to match the observed levels (white). Note the difference in velocity dispersion between the
 Z = 0$\arcsec$ PVDs at 3 and 9$\sigma$. Although the modeling includes the lognorm radial inflow discussed in the text, note that the modeled PVDs
 do not reproduce any ``beard'' feature.}
\end{center}
%

%
 \noindent
\includegraphics[scale=0.9]{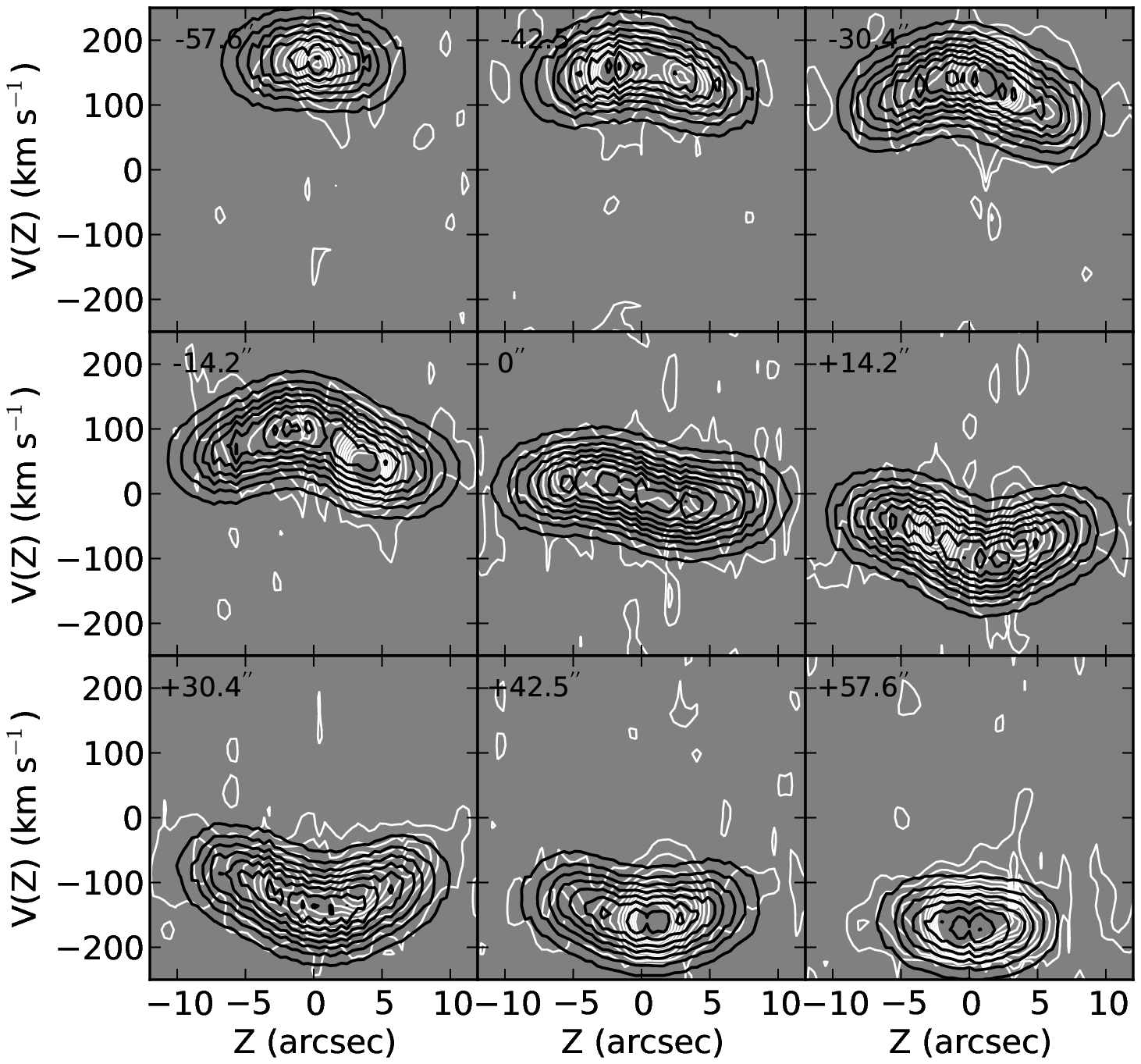}
\figcaption{\label{fig. 13} Modeled PVDs parallel to minor axis including lognormal inflow and an exponential plus ring radial
 distribution(Model R, black contours).
 Contour levels are in 3$\sigma$'s of emission signal normalized to match the observed levels (white).
 The addition of an inflow reproduces satisfactorily the kinematic asymmetry displayed in Figure 10.}
%

\begin{figure}
\centering
\includegraphics[width=\textwidth]{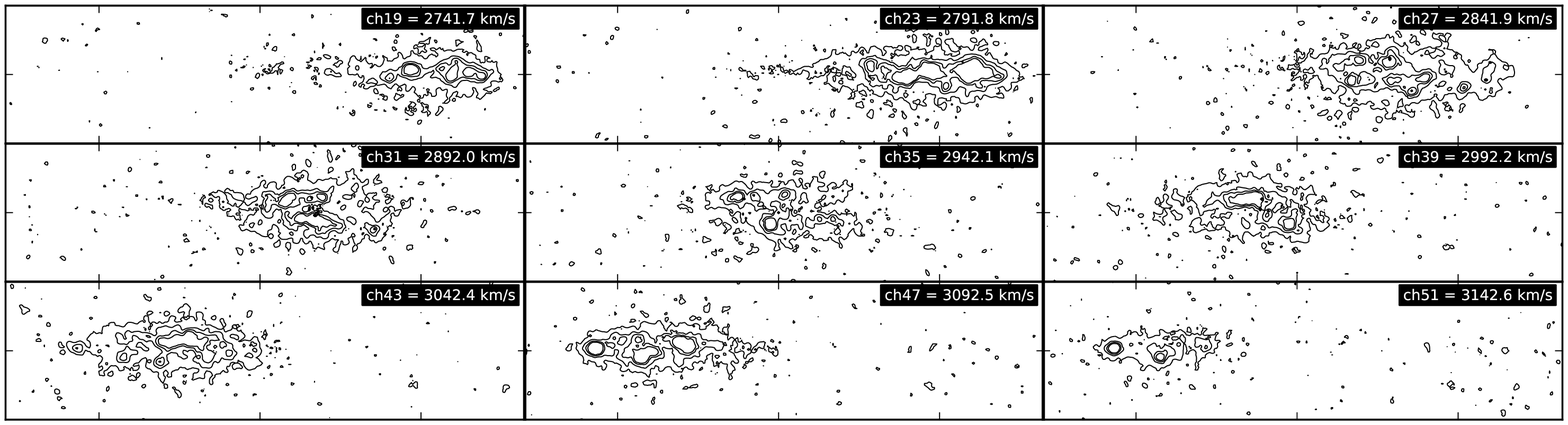}
\vskip 0.2in
\centering
\includegraphics[width=\textwidth]{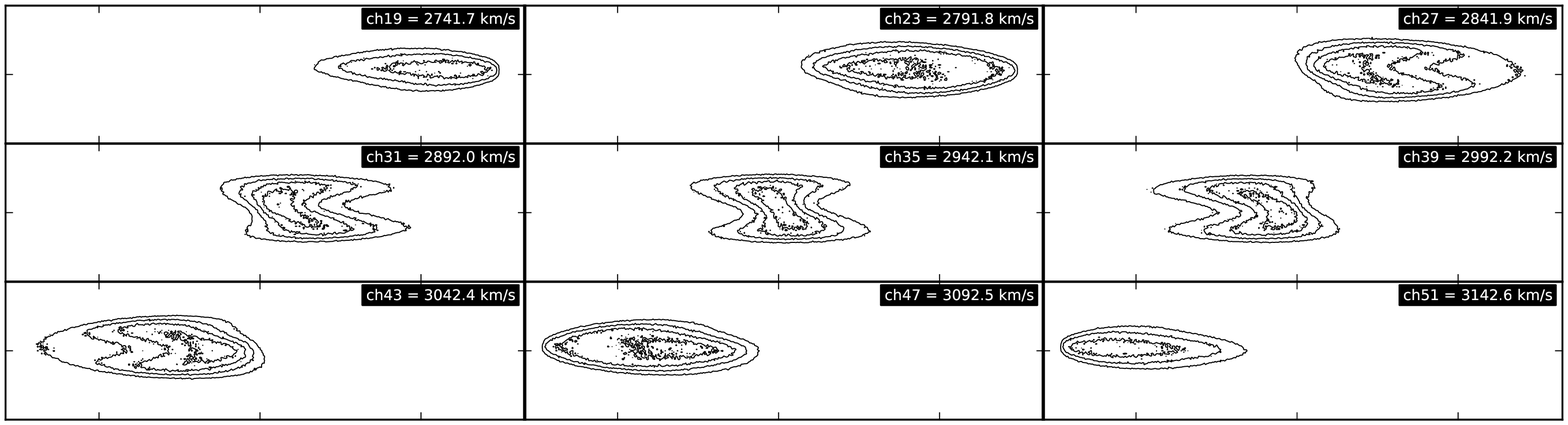}
\vskip 0.2in
\centering
\includegraphics[width=\textwidth]{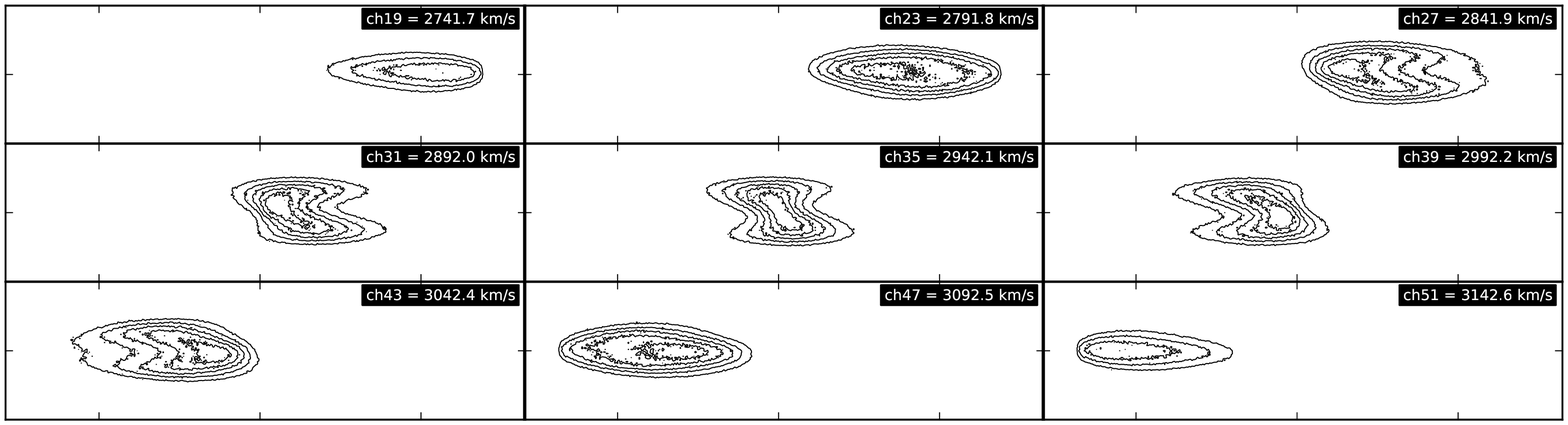}
\figcaption{\label{fig. 14} Observed (top) and modeled (medium: Model WR and bottom: Model R) channel maps. Contour levels are in
 $3$, $6$, $9$, and $12\sigma$'s of emission signal. For the modeled channel maps the levels are normalized
 to match the observed ones. Horizontal and vertical tick separations are
$47\farcs8$ and $20\farcs2$, respectively. Note the bifurcation in the observed velocity channels depicted in the central panel. Note that the bifurcation appears even without considering the ring of HII regions of Model R.}
\end{figure}

\subsubsection{ Determining the inclination }

In Section 3 we have made an estimate of the inclination  ($\it i$ =
 82$\arcdeg$.5) based only on morphological grounds.
In order to know better this parameter, we have run  models with different
 inclinations, radial distributions
 and vertical scaleheight values in order to see whether some combination
of these parameters reproduces the observed maps. The results are as follows:

The synthetic velocity maps obtained from models with assumed inclinations of
 85$\arcdeg$ and 88$\arcdeg$ and an exponential radial distribution present
 several problems: 1) They cannot reproduce the emission at Z =$\pm$
 8$\arcsec$.1 seen in Fig. 8 even if we increase the vertical scaleheights
 values to 
up to 5$\arcsec$,  2) while for $\it i$ = 85$\arcdeg$ the synthetic velocity
 maps show the bifurcation found in the velocity maps 
derived from
our kinematical observations, corresponding to the central panels of Fig. 4
 (although it is roundish for high vertical scaleheighs needed to fulfill point
1), for  $\it i$ = 88$\arcdeg$ no bifurcation in the velocity maps is
 reproduced. Models with
 smaller values of the inclination like $\it i$ = 80$\arcdeg$ predict more
 pronounced bifurcations than what is seen in the velocity maps 
derived from
our kinematical observations.

For the PVDs parallel to the minor axis, there are also discrepancies between
the observed PVDs and the PVDs derived from models with larger or smaller 
values of the
inclination than the adopted value. Indeed, for smaller values of the
 inclination (i.e., $\it i$ = 80$\arcdeg$), the PVDs parallel to the minor axis
 appear too extended, showing emission well outside the observed contours. On
 the other hand, for models with inclination values of 85$\arcdeg$ and  
88$\arcdeg$  (an almost edge-on galaxy), the PVDs parallel to the minor axis
 show a roundish appearance, different to the observed heart-like PVDs (see
 Figure 10).

We also explored how the different synthetic maps change if we include in 
the radial distribution besides an exponential disk with 
a scale radius of 200$\arcsec$ 
an axisymmetric ring such that the
total surface density is $\Sigma(R)=\Sigma_0\exp(R/h)+\Sigma_{ring}(R)$. The
 ring is simulated as a Gaussian function:

\begin{equation}
\Sigma_{ring}(R)=A_r\Sigma_0\exp\left(-\frac{(R-\mu)^2}{2s^2}\right)
\label{Gauss_ring}
\end{equation}

We first tried the parameters $\mu=24\arcsec$ and $s=10\arcsec$, selected to roughly
 match the location of the peak and the width of a lognormal radial velocity 
inflow which is required to fit the observed radial velocity field and PVDs
parallel to the minor axis (see Sect. 4.3.3). The amplitude of this ring
was chosen as $A_r=0.4$ as a suitable value to take into account the ring of
HII regions noticed in Sect. 3.  We have found that introducing a ring in the brightness
 distribution, with those parameter values, makes that a bifurcation in the central velocity maps appears for $\it i$ = 85$\arcdeg$ (while no bifurcation appears in the velocity maps for
 $\it i$ = 88$\arcdeg$). However, the bifurcation is not as pronounced as the
 observed one, even changing the ring amplitude (we have considered other amplitude values such as 1.4 and 8.0). Figure 14 shows this and Figure 15 shows the effect of considering that ring of HII regions in the synthetic PVDs parallel to the minor axis. On the other hand, the PVDs 
at Z =$\pm$ 8$\arcsec$.1 seen in 
Fig. 8 are still not reproduced for $\it i$ = 85$\arcdeg$, even if we increase
 the  
vertical scaleheight up to Z = 5$\arcsec$.

We also tried a more extreme ring distribution by fixing the ring parameters $\mu=60\arcsec$ and $A_r=8.6$. In the case of $\it i$ = 85$\arcdeg$, we recover the bifurcation of the central velocity maps, but still, it is not possible to fit the  PVDs 
at Z =$\pm$ 8$\arcsec$.1. For $\it i$ = 87$\arcdeg$, we cannot reproduce the bifurcation of the central velocity maps, nor the emission at  Z =$\pm$ 8$\arcsec$.1; besides, the minor axis PVDs are not well reproduced.

We have also varied the vertical scale height from its adopted value of
 2$\arcsec$ running models with larger (5$\arcsec$) and smaller (1$\arcsec$.5)
 values. We find that the models with the adopted value reproduce better the
 observations.

\subsubsection{ A plausible (best) fit }

Taking into account the arguments given above, we adopted an inclination 
of $\it i$ = 82$\arcdeg$.5, based both on the
 morphology of the galaxy and in the fitting to the kinematical observables we
 have such as velocity maps and PVDs. This value could be in error by 1 to 2
 degrees. We have also adopted a vertical scaleheight of 2$\arcsec$, a radial
distribution consisting of an exponential disk of 200$\arcsec$ scale radius
with a ring of HII regions described by Eq. (5) with $\mu=24\arcsec$ and $A_r=0.4$, with velocities composed
of two parts: the circular velocities given by the ET RC and a lognorm
radial inflow described by Eq. (4). The disk is modeled as a series of
annulii having all of them the same inclination and position angle of
67$\arcdeg$ and the systemic velocity and center of rotation are given in 
Table 2. Figures 12, 13 and 14 show the comparison between the observed
PVDs parallel to the major axis, parallel to the minor axis and the velocity
maps and  the corresponding synthetic ones derived from the GALMOD model
we selected. As it can be seen, the features listed in 4.3.2 are reasonably
reproduced.

The model with the values for the different parameters listed above is called Model R. Most of the comparisons between the observed data and the synthetic
modeled data are done considering Model R. However, we illustrate in some cases the behaviour of some parameters by showing in some figures the synthetic
modeling of other models that have the same parameters as Model R but either
they have only an exponential radial distribution (without ring): Model WR, or 
they do not consider any inflow: Model WI.

 \noindent
\includegraphics[scale=0.9]{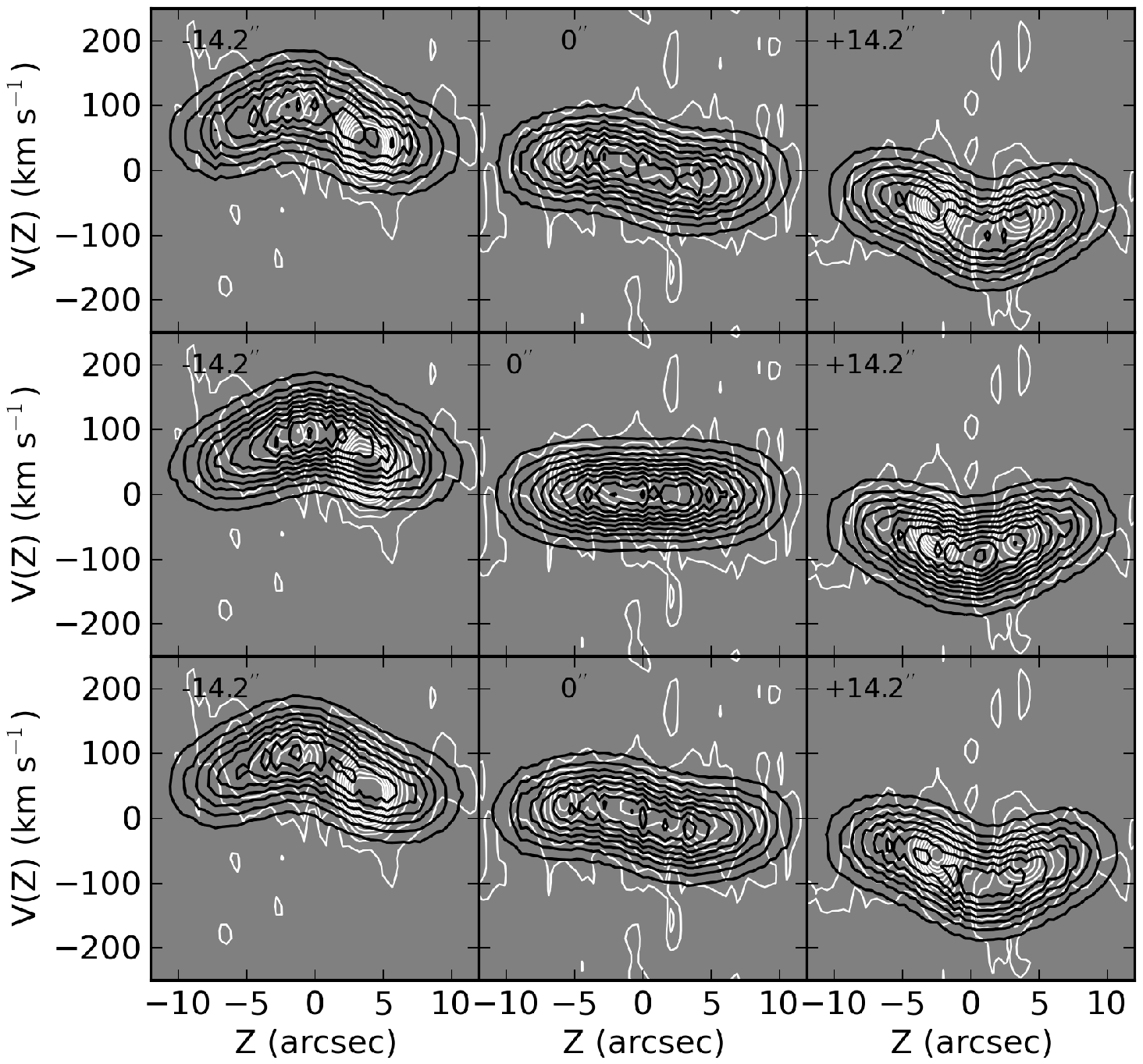}
\figcaption{\label{fig. 15} Comparison of the observed (white contours) and modeled (black contours) PVDs parallel to the minor axis for three different models. Upper panel: the model used here  considering lognorm inflow but without a ring (Model WR, see the text). Middle panel: the model used here, considering a ring but without inflow (Model WI). Lower panel: the model used here, considering both the ring and lognorm inflow (Model R).}
%

\subsection{ Dust Extinction}

In addition to LOS effects discussed before, dust extinction in the disk has been invoked to preclude the derivation of RCs that really map the gravitational potential of a galaxy in the case of the inner parts of edge-on galaxies. ESO~379-006 does not show a clear dust lane at the midplane as revealed by the unsharp-masked R image published in Rossa \& Dettmar (2003b). On the other hand, Mathews \& Wood (2001) have used a combination of 3D Monte Carlo radiative transfer techniques and multi-wavelength imaging data to investigate the nature of the interstellar medium (ISM) in an edge-on galaxy of low surface brightness (LSB). They have also used realistic models that incorporate multiple scattering effects and clumping of the stars and ISM (gas and dust) and its effects on the observed stellar disk luminosity profiles, color gradients and RC shape. In what concerns the RC, in spite of the clear presence of dust in the galaxy studied, the authors conclude that realistic optical depth effects will have little impact in the observed RCs of edge-on disk galaxies and cannot explain the linear, slowly rising RCs seen in some edge-on LSB spirals. They also concluded that projection effects (as the LOS effects studied by us before) create a far larger uncertainty in recovering the true underlying RC shape of galaxies viewed at large inclinations.

\section{Discussion and conclusions}  

We have seen that in order to fit the  asymmetry in velocities of the observed PVDs parallel to the minor axis and to fit the skewed iso-velocity contours in the observed  velocity field with the galaxy model we used, we need to include, in addition to a disk causing the rotation motion of the gas, an inflow motion in the radial direction. We have shown that it is indeed an inflow and not an outflow given that we were able to determine the nearest side of the galaxy. We tried different laws for the inflowing gas and we have found that the lognorm law fitted better the observations. This implies that the inflow velocity is nearly zero  at the center (within a radius $\sim 15\arcsec$, or 2.7 kpc), reaches a maximum velocity of about 25 km s$^{-1}$ at $\sim 20\arcsec$ (equivalent to 3.6 kpc) and afterwards it decays with R. The question now is to distinguish the mechanism that could cause the inflow. 

To find non-circular motions that can be modeled some way with an inflowing motion (even if it is of the ``weird'' lognorm form) and to have skewed iso-velocity
 contours in the velocity field of ESO~379--006, suggest the existence of a bar. It is difficult to say more about the specific characteristics of this hypothetical bar because we have only kinematical traces of it and it is impossible
to confront with any morphological data 
 because the inclination of this galaxy precludes to have a good view of  the extension, strength and alignment of that  bar. 

In that context, the lognorm inflow that fits the best our kinematic data has a maximum velocity of 25 km s$^{-1}$ between 20$\arcsec$ to 40$\arcsec$ (for radii smaller or larger than this value the inflow velocity is much smaller),
suggesting a resemblance with the velocity jump of tens of km s$^{-1}$ that Sofue \& Rubin (2001) identify as characteristic on the 
leading edges of bars. Furthermore,  skewed  velocity contours  similar as the one we derived in this work has been found in less inclined barred-galaxies (see for example Duval et al. 1991, Sofue \& Rubin 2001, and references therein).

According to Sparke \& Gallagher (2000; see also Binney \& Tremaine 1987), in a bar, the stars rotate in closed orbits that are  very elongated and aligned with the bar axis. The kinematics of gas within the
bar is such that gas stays close to the elongated stellar orbits,
till the end of the bar (presumably corotation lies beyond the bar's end)
where the gas flow converges, gas pressure rises and a shock forms.
In the shock the gas dissipate its kinetic energy as it heats itself and thus,
it starts to drop onto  orbits closer to the center of the galaxy. This
inflow continues until the gas reaches the very interior rounder orbits 
lying perpendicular to the main 
bar axis. Then, the inflow stops and the gas piles up
forming a central ring. 
We cannot assure that this is happening in ESO~379--006 but it is a plausible
scenario that can explain the observed kinematics of this galaxy.

Another possibility is the influence of spiral arms. Indeed, a grand-design spiral pattern could induce  non-circular motions via either an hydraulic jump considering either a bar or a spiral density wave as obstacles (Martos \& Cox 1998) or just only by the presence of the asymmetric potential of the spiral arms. Besides, the presence of active star formation can trigger galactic winds and superbubble expansions that could be seen as non-circular motions.  
This galaxy certainly has spiral arms. In this particular case we have some evidence of the presence of those spiral arms as we noticed in Section 4.1 (see Figures 1 and 6).  The gas motions  both in the spiral arms and in the inter-arm regions could cause non-circular motions as well, not necessarily as an inflow. We should remind from the Introduction that ESO~379--006 is not an starburst galaxy; it has an SFR comparable to the SFR of NGC~891 while the energy input from SN explosions per unit area is 10 times smaller than the value for NGC~891. If NGC~891 is considered as the prototypical galaxy where extended e-DIG is detected (perhaps caused by mildly galactic winds), ESO~379--006 shows much fainter filamentary  e-DIG  spreading in different locations over the whole extension of the disk, as we have noticed in Sections 3 and 4.2.1. However, it is possible that both the spiral arms as obstacles and as sites of SF in the galaxy contribute in a combined action to generate the non-circular motions modeled by us as an inflow.

In principle, the induced non-circular motions are larger for bar-induced motions (about 40 to 20 km s$^{-1}$) than for perturbations due to spiral arms (about 20 to 10 km s$^{-1}$;  Sofue \& Rubin, 2001). The inflow we modeled here gives maximum velocity values of 25 km s$^{-1}$ (which, we have shown this inflow velocity is linked to the value of the observed PVDs kinematic asymmetry) thus just in between both possibilities (favoring slightly the bar).

We started this kinematical study with the aim of knowing more about the kinematics of the e-DIG in this galaxy. We indeed detect from our images this faint component, as previously other authors did. Furthermore, we detect  faint gas with velocities different to the rotation motion of the disk (we favored ``the beard'', although we cannot discard LOS projection effects, Section 4.2.1). The detection is clearer for the approaching side of the disk while for the receding side  the possible detection of the faint gas is contaminated by the [NII] line emission of the disk. Following the interpretation of similar results in the PVDs of HI emission of less inclined galaxies, we suggest that this kinematically different, faint component corresponds to e-DIG lagging in rotation relative to the disk. However, given the low intensity of the component  we did not get enough signal in the PVDs parallel to the major axis  to model this faint component as an additional motion in Equation 3 (i.e., by adding a velocity component along the Z axis, or even, by adding another component to the rotation motion). Consequently, we were unable to determine its height above or below the disk or a possible rotation velocity gradient with Z. Deeper kinematical observations are required in order to confirm these results and to know better the velocity law of the faint gas along the Z-height.

The existence of e-DIG in this galaxy cannot help us in discriminating between the possibilities mentioned above as causing the lognorm radial inflow (a bar or spiral arms) because both agents could produce the escape of gas from the disk. For instance, the existence of a bar could favor vertical motions (along the Z-axis) of the gas in the disk as it encounters an obstacle like a bar (i.e., an hydraulic jump according with Martos \& Cox 1998 and Martos et al. 2012). On the other hand, spiral arms, by means of the SF and SN activity are capable of forming expanding superbubbles and also spiral arms could act as obstacles triggering hydraulic jumps that can provide  a velocity component in the vertical direction.

Thus, we can conclude the following:

 In this work we have studied the morphology and kinematics of the ionized gas in the 
nearly edge-on galaxy ESO~379--006 by means of Fabry-Perot spectroscopy.
 We succeeded in obtaining rotation curves of the main disk of the galaxy by means 
of the Intensity-Peak (barycentric) and ET methods. The maximum amplitudes 
derived from the RCs are quoted in Table 2.

{\it i})  Our modeling shows that the Envelope-Tracing RC
fits better our PVD along the major axis than the barycentric RC, indicating that projection effects are present for this nearly edge-on galaxy. The ET RC gives a maximum amplitude 
of 200 km s$^{-1}$, slightly larger than the amplitude reported in previous works.

{\it ii}) We also have found a velocity asymmetry in the PVDs along or parallel to the minor axis  
corresponding to central regions. The deviations to circular motion in the central parts
of this galaxy together with the skewed isovelocity contours found in the radial 
velocity field suggested to us that another velocity component different to the rotation was required to model the observed motion of this galaxy.

Thus, we have added to the motion of a thin disk with circular rotation given by the Envelope-Tracing RC, a radial inflow of the gas (lognorm) of 
about 25 km s$^{-1}$ in its maximum value.  We have also varied the inclination of the galaxy and the vertical scale height in order to select the values that better fit the data, obtaining an inclination of 82$\arcdeg$.5 and a scale height of 2$\arcsec$.
With this additional motion, and the values of inclination and scale height given above, the confrontation between models and  data extracted from observations is  better.

{\it iii}) The existence of non-circular motions modeled as a radial inflow
could be due to several possibilities. We have examined only two of them: a bar, and the spiral arms with SF activity. We found that there is no way with the present data to distinguish between those possibilities.

{\it iv}) We suggest that either the existence of a bar or the presence of the spiral arms could explain both the particular radial inflow and
the existence of e-DIG. However, the inclination of this galaxy does not allow us to discriminate between those possibilities.

{\it v}) We detected e-DIG in the form of filaments and other very local patches sorting out 
from the disk. However, no diffuse, generalized emission spread over the entire 
disk has been detected in the H$\alpha$ images we have built from the FP data cubes.

{\it vi}) In addition, in a surprising way, we have marginally detected e-DIG by means of its different 
kinematical behavior relative to the main disk rotation. The e-DIG identified this way 
is faint and following one possible interpretation, it seems to lag in velocity relative to the corresponding circular velocity 
of the gas in the disk. However, from the way we have detected it, it is not possible to
know at 
which vertical height the gas is.

\acknowledgments

The authors wish to thank Mrs. Jana Benda for reading the manuscript and to Carmelo Guzm\'an for helping in computer facilities. We thank the referee for helping us to improve the paper. This paper was done with financial support from grants  P-82389 from CONACYT and  grants IN102309 and IN108912 from DGAPA-UNAM. IFC acknowledges grants 0133520 from CONACYT and 20121700 from IPN.

\clearpage

\begin{deluxetable}{lcc}
\scriptsize
\tablewidth{0pc}
\tablenum{1}
\tablecaption{Main Properties of the Galaxy ESO 379-006.}
\tablehead{ Property &   & References or Notes \\
   }

\startdata

Name            &  ESO379-006 &  \\
Right Ascension & 11$^h$ 53$^m$ 2$^s$.4 & J2000.0   \\
Declination     & -36$\arcdeg$ 38$\arcmin$ 20$\arcsec$.0 & J2000.0 \\
D$_{25}$        & 2$\arcmin$.75/31.82 kpc  & LEDA  \\
Distance        & 38$\pm$3 Mpc & RD\tablenotemark{(1)}  \\
V$_{helio}$     & +2940 kms$^{-1}$ & NED  \\
Type            &   Sc   &   NED    \\
m$_B$           & 14.13  &   NED    \\
S$_{60}$/S$_{100}$      & 0.2452   &  RD\tablenotemark{(1)} \\
L$_{FIR}$/$(D_{25})^2$  & $3.67 \times 10^{40}$ erg s$^{-1}$ kpc$^{-2}$  & RD\tablenotemark{(1)}  \\
SFR$_{FIR}$ & 1.49 M$_\odot$ yr$^{-1}$ & RD\tablenotemark{(1)}  \\
$\nu_{SN}$  &  0.061 yr$^{-1}$   &  RD\tablenotemark{(1)}  \\
dE$_{SN}$/dt $\times$ Area$_{SF}^{-1}$ & 4.1 $\times 10^{-4}$ erg s$^{-1}$ cm$^{-2}$  &  RD\tablenotemark{(1)}  \\

\enddata
\tablecomments{(1):  from Rossa \& Dettmar 2003a }
\end{deluxetable}

\begin{deluxetable}{lcccc}
\scriptsize \tablewidth{0pc} \tablenum{2} \tablecaption{Morphological and Kinematical
Properties of the Galaxy ESO 379-006 and its e-DIG.} \tablehead{ Property &
MFB\tablenotemark{(1)} & PS\tablenotemark{(2)} &
RD\tablenotemark{(3)} & This work \\} \startdata
Type & Sc & -- & Sbc & SBc  \\

2a $\times$ 2b & 2$\arcmin$.82 $\times$ 0$\arcmin$.5 & 2$\arcmin$.3 & 2$\arcmin$.6 $\times$ 0$\arcmin$.4 &  3$\arcmin$.02 $\times$ 0$\arcmin$.4\tablenotemark{(4)}\\
i & 90$\arcdeg$ & 90$\arcdeg$ & 81$\arcdeg$ & 82.5$\arcdeg$  \\
PA  & 68$\arcdeg$  &  -- & -- & 67$\arcdeg$ \\
V$_{helio}$ & +2940 kms$^{-1}$ & +2944 kms$^{-1}$ & +2986 kms$^{-1}$\tablenotemark{(5)} & +2944 $\pm$3 kms$^{-1}$  \\
V$_{ROT-MAX}$ & 171 kms$^{-1}$  &  169 kms$^{-1}$ & -- & 200 kms$^{-1}$\tablenotemark{(6)} \\
V$_{ROT-MAX}$ &  &   &  & 200 kms$^{-1}$\tablenotemark{(7)} \\

offset of kinematical center& 0$\arcsec$ & 2$\arcsec$.3 & --  & 1$\arcsec$.8 \\
from photometric center  &   &   &  & to the NE\\
$\sigma_{LOS}$ for H$\alpha$ & -- & -- & -- & 35--40 kms$^{-1}$ \\
Maximum inflow velocity V$_0$ & -- & -- & -- & 25  kms$^{-1}$ \\

Z$_{MAX}$\tablenotemark{(8)}  &  -- &  -- &  2.08 kpc & 1.84 kpc\\
\enddata
\tablecomments{(1): MFB: Mathewson, Ford \& Buchhorn 1992;~ (2): PS: Persic \& Salucci 1995;~ (3): RD: Rossa \& Dettmar 2003;~ (4): Dimensions adopted from NED;~(5): RC3 (de Vaucouleurs et al. 1991);~(6): From ET Method RC;~(7): From Intensity-Peak (Barycentric) RC;~(8): Maximum height above or below the disk of the e-DIG, estimated from the angular distance to the major axis.}
\end{deluxetable}

\clearpage

\end{document}